\documentclass[a4paper,11pt]{article}
\pdfoutput=1 

\usepackage{jheppub} 

\usepackage[T1]{fontenc} 
\usepackage{caption}
\usepackage{tikz}
\usepackage{subcaption}
\usepackage{graphicx}
\usepackage{array}
\usepackage{makecell}
\usepackage{longtable}
\usepackage{float}
\usepackage{bm}
\usepackage[export]{adjustbox}[2011/08/13]

\definecolor{darkgreen}{rgb}{0.1,0.5,0.1}
\definecolor{cerulean}{rgb}{0.0, 0.48, 0.65}

\newcommand{\cmssm}{cMSSM}
\newcommand{\pmssm}{pMSSM}

\usepackage{amsmath}
\usepackage{graphicx}

\title{Visualization and Efficient Generation of Constrained High-dimensional Theoretical Parameter Spaces}
\author{Jason Baretz$^{a}$}
\author{Nicholas Carrara$^{b}$}
\author{Jacob Hollingsworth$^{a}$}
\author{Daniel Whiteson$^{a}$}

\affiliation{$^{a}$Department of Physics and Astronomy, University of California, Irvine, CA, USA 92627}
\affiliation{$^{b}$Department of Physics, University of California, Davis, CA, USA}

\abstract{ We describe a set of novel methods for efficiently sampling high-dimensional parameter spaces of physical theories defined at high energies, but constrained by experimental measurements made at lower energies. Often, theoretical models such as supersymmetry are defined by many parameters, $\mathcal{O}(10-100)$, expressed at high energies, while relevant experimental constraints are often defined at much lower energies, preventing them from directly ruling out portions of the space.  Instead, the low-energy constraints define a complex, potentially non-contiguous subspace of the theory parameters. Naive scanning of the theory space for points which satisfy the low-energy constraints is hopelessly inefficient due to the high dimensionality, and the inverse problem is considered intractable. As a result, many theoretical spaces remain under-explored. We introduce a class of modified generative autoencoders, which attack this problem by mapping the high-dimensional parameter space to a structured low-dimensional latent space, allowing for easy visualization and efficient generation of theory points which satisfy experimental constraints. An extension without dimensional compression, which focuses on limiting potential information loss, is also introduced. 

}

\begin{document}
\maketitle

\section{Introduction}

 Decades of searches for extensions to the Standard Model such as supersymmetry (SUSY) have come up empty, and future higher-energy colliders may be required in order to discover new particles and interactions.  But given the vast expense and wait time incurred by such projects, it is worth exploring diligently whether the parameter spaces of these theories have been thoroughly exhausted in current datasets. Could there still be novel unexplored islands\textemdash unexcluded regions of parameter space with signatures accessible to the LHC and astroparticle experiments? 

The full parameter space of the minimal super-symmetric model (MSSM) has more than 100 parameters~\cite{MARTIN_1998}, making a complete exploration daunting. Out of desperate pragmatism, the search space is often restricted to a more tractable subset of parameters based on theoretical inclinations. Common examples include the ($4+1$)--dimensional  {\cmssm} as well as the 19--dimensional  {\pmssm}.  These  high-dimensional spaces guide the LHC program ~\cite{PhysRevD.86.055007, HAN2017470,PhysRevD.83.095019, Buchmueller2014,Bridges:2011,https://doi.org/10.48550/arxiv.1307.8444,PhysRevD.91.055002,Aad_2015,CMS:2016}, but direct experimental searches are typically performed in only one or two dimensions, with all other parameters fixed to theoretically preferred values. The resulting surfaces of LHC exclusion occupy a vanishingly small fraction of the volume of the larger theoretical space, most of which is essentially unexamined. Each point in such a space defines a spectrum of particles which determine well-known quantities such as the Higgs boson mass and the dark matter relic density.  But what portions of the full SUSY space are consistent with these weak-scale constraints and LHC exclusion results? Are there unexplored islands where promising models abound? Nobody knows.

A critical obstacle is that weak-scale constraints such as the Higgs mass or the dark matter relic density are not expressed directly in terms of the fundamental parameters of the theory, such as SUSY particle masses, instead requiring intensive calculations. This prevents us from using experimental constraints to immediately rule out large portions of the model's parameter space or reduce its dimensionality. Instead, the weak-scale constraints define a complex, potentially non-contiguous subspace of the theory parameters. And the inverse problem, determining a map from weak-scale observables to fundamental parameters, is intractable. Instead, we are forced to scan the high-dimensional parameter space, evaluating each point for weak-scale consistency. Efforts to speed up the calculation of the weak-scale quantities with machine learning ~\cite{Caron2017, Kronheim_2021, Bridges_2011} 
do not solve the core problem, which is dominated by the dependence of the scanning cost on the dimensionality rather than the per-sample computation time. 

Even the reduced spaces such as the {\pmssm} are difficult to scan using a brute-force search, so much so that the discovery of new islands of experimental consistency in this subspace relies largely on the inspiration of particle theorists intuiting their existence. Each island (e.g. ``well-tempered SUSY''~\cite{Cheung:2012qy}, ``focus point SUSY''~\cite{Feng:1999zg}, ``Natural SUSY''~\cite{Papucci:2011wy}, ``Gauge Mediated SUSY Breaking'' ~\cite{Dine_1993}, ``General Gauge Mediation''~\cite{Meade_2009}) is then carefully studied~\cite{Kowalska2013NaturalMA,Hall2011ANS,Huang2014BlindSF,Feng_2012,Kowalska2014LowFT,Evans_2014,Buckley_2017_1,Buckley_2017_2,Dine_1995,Dine_1996}, and novel signatures can motivate new experimental analyses. But the danger of this exploration by intuition is that one may miss viable regions. Currently, many particle physics theories include vast, mostly-unexplored parameter spaces, which may contain undiscovered islands consistent with experimental constraints. Some of these islands may reveal new unanticipated signatures accessible at the LHC.
 
Fortunately, an arsenal of techniques in Artificial Intelligence (AI), and in particular statistical techniques such as Machine Learning (ML), have shown great promise at exploring and summarizing information from high-dimensional data spaces.  Recent approaches for exploring high dimensional HEP spaces fall largely into two broad categories.

 A class of generative method such as GANs \cite{goodfellow2014generative}, HMC \cite{betancourt2018conceptual,neal2012mcmc,Baltz_2004}, Normalizing Flows \cite{Kobyzev_2021}, and Genetic Algorithms \cite{Goldberg1988GeneticAI}, take advantage of the ability of ML to perform high-dimensional interpolation, making  notable improvements in the efficiency of sampling high-dimensional subspaces consistent with experimental constraints ~\cite{Hollingsworth:2021sii, Morrison:2022, Abel_2014,Souza, https://doi.org/10.48550/arxiv.2207.05103,https://doi.org/10.48550/arxiv.1805.03615}.  But these methods often act as black boxes, making interpretability difficult, and ameliorate but do not fundamentally solve the underlying difficulty of searching high-dimensional spaces. The increasingly high-dimensionality of SUSY-related theoretical spaces may instead require methods which map the problem to a lower-dimensional space where extensive sampling is more feasible.

The other broad category of approaches, observational dimensional reduction methods (e.g. UMAP \cite{https://doi.org/10.48550/arxiv.1802.03426}, TSNE \cite{JMLR:v9:vandermaaten08a}, non-generative auto-encoders \cite{ACKLEY1985147}, and self-organizing maps \cite{2007SchpJ...2.1568K,Kohonen2004SelforganizedFO}), analyze high dimensional datasets by mapping them to a lower-dimensional latent space~\cite{Mutter:2019,he2022machinelearning}, allowing for {\it visualization} of any resulting structure, such as fertile islands of desirable points. For example, in the context of SUSY parameter space searches, these techniques allow for visual examination of the latent space in the hopes of finding clusters of points consistent with experimental constraints. While the goal of collapsing the high-dimensional structure into a lower dimensional space has solid mathematical underpinnings~\cite{Fefferman_Mitter_Narayanan_2016}, actually creating useful structure is challenging, as many of the methods are unsupervised and thus cannot explicitly encourage such structure in the latent space. Often, the resulting latent space is too poorly structured for efficient sampling ~\cite{Kolouri:2018}. In other cases, the lack of a return mapping from the latent space prevents sampling altogether. However, when used in the context of generative methods, reducing the dimensional complexity of the sampling space has the potential to directly remove the primary difficulty posed by searching high-dimensional spaces.

Additionally, Manifold Learning methods ~\cite{park2022neural,brehmer2020flows,kumar2022manifold,kumar2022shining} spanning both of these categories have sought to learn and sample from low-dimensional manifolds on which interesting points sit in high dimensional spaces, although they can be non-trivial and computationally expensive to train ~\cite{park2022neural,brehmer2020flows}.

\begin{figure}[H]
\centering
\includegraphics[scale=0.6]{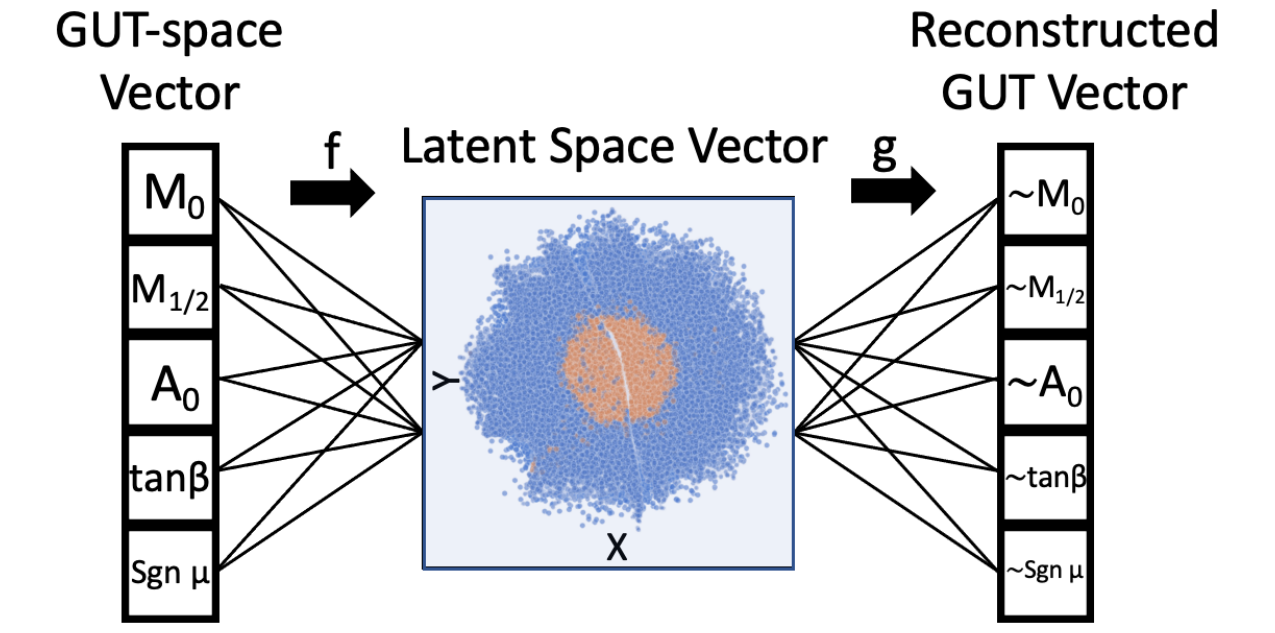}
\caption{ A simplified diagram of SAFESPAM for the \cmssm. GUT-scale inputs are mapped to an abstract 2D latent space, where they are shaped to a target distribution via loss function. Weak scale values that accompany each theory point are used only to classify points as valid (orange) or invalid (blue), and do not enter the network as inputs. Valid points shown in orange cluster towards the center of the latent space. The mapping $f$ from input theory space to latent 2D space is referred to as the {\it encoder} and the return mapping $g$ as the {\it decoder}. SAFESPAM tries to reconstruct the GUT scale parameters as accurately as possible by minimizing the L2 term in the loss function, ensuring that the outputs from the decoder match the corresponding inputs to the encoder. Sampling new points in the valid region of the latent space allows for efficient generation of new valid points in GUT space.}
\label{fig:sampex}
\end{figure}

 In this paper, we build a new class of generative methods which map the problem to a lower-dimensional space, but specifically encourage useful structure by treating the problem as a supervised learning task.  We use a modified Sliced Wasserstein Autoencoder (SWAE) ~\cite{Kolouri:2018} structure to create bidirectional maps between the GUT-scale parameter space and a {\it structured low-dimensional latent space}. The structure of the latent space is specialized to force valid points to cluster near the latent origin, allowing for efficient generation of new points which can be mapped back to the GUT-scale parameter space. Because the latent space is low-dimensional, it is inherently easier to search. Additionally, the added benefit of visualization due to dimensional reduction makes our maps easy to understand and interact with, and in some cases allows the addition of secondary constraints {\it in media res} with no additional training. Because our method reduces the complexity of the sampling space, it may be scalable in the future to tackle larger, less-constrained subsections of the MSSM. The structure of our approach shown in Fig.~\ref{fig:sampex}. To explore whether encouraging structure in the latent space is sufficient, without insisting on dimensional reduction,  we also introduce a complementary formulation of our method which operates without any dimensional reduction. 
 
We call our primary method the SWAE Algorithm For Experimentally Sound Point-Production And Mapping (SAFESPAM), and two variants without dimensional reduction we refer to as Constraint-driven High-dimensional UNcompressed (Categorical) Clustering, or CHUNC and CHUNC2\footnote{\url{https://github.com/infophysics/CHUNCNet}.} \cite{CHUNCNet}. 

The paper is outlined as follows: In Section II we detail the task and nature of our experimental constraints. Section III describes the generation and parameters of our dataset. In Section IV, we discuss our approach for 
SAFESPAM, CHUNC and CHUNC2, as well as our sampling procedures. Section V presents the results of our trials for various sets of experimental constraints, and Section VI contains further discussion and conclusions.

\section{Task}

To efficiently generate new points in high-dimensional theoretical parameter spaces which satisfy the experimental constraints placed on weak-scale observables, we seek maps between the theory space and a structured latent space which are:

\begin{itemize}
    \item {\it generative}: capable of generating new points with high efficiency compared with naive brute force sampling
    \item {\it contrastive}: with structured latent spaces where valid and invalid points are separated into distinct regions
    \item {\it low-dimensional}: for ease of sampling and visualization of the latent space
\end{itemize}

In this paper, we study the performance of the SAFESPAM method in two constrained SUSY spaces, the {\cmssm} in 4+1 dimensions and {\pmssm} in 19 dimensions ~\cite{Cohen:2013,Djouadi:1999,PhysRevLett.49.970}. We impose two experimental constraints, the Higgs boson mass and the dark matter relic density:   
\begin{itemize}
    \item  Higgs Boson Mass: 122.09 GeV $\leq m_h \leq $ 128.09 GeV ~\cite{20121,ce431d39f29d462482f95f2cb27c9583}
    \item Dark Matter Relic Density:  0.08 $\leq \Omega_{DM}h^2 \leq $ 0.14  ~\cite{Spergel_2003,Bennett_2013}
     \end{itemize}
     
Points are denoted as `valid' whose experimental values fall within these constraints. While in principle our methods could also use other experimental variables as constraints, prior work has often focused on these ~\cite{Hollingsworth:2021sii,Morrison:2022,Souza}. With CHUNC and CHUNC2, we introduce a third constraint that the Lightest Supersymmetric Particle (LSP) be a neutralino. All weak scale constraints are calculated using softsusy 4.1.10 and micromegas 5.2.6~\cite{ALLANACH2002305}~\cite{micrOMEGAs}. 

 \par

\section{Dataset}

  We utilize initial brute-force scans, $\mathcal{O}$(10-20M) points, of the parameters of {\pmssm} and {\cmssm} at the GUT scale to create our datasets ~\cite{Cohen:2013,https://doi.org/10.48550/arxiv.1307.8444}. Below we denote the  parameter ranges for the \cmssm~(Table~\ref{tab:cmssm}) and \pmssm~(Table~\ref{tab:pmssm}). All parameters are sampled uniformly over the specified range, with the exception of $A_0$, where $\frac{A_0}{m_0}$ is sampled uniformly ~\cite{Hollingsworth:2021sii}.  

GUT-scale points are considered to be {\it valid} if they are consistent with experimental constraints, calculated via softsusy 4.1.10 and micromegas 5.2.6~\cite{ALLANACH2002305}~\cite{micrOMEGAs}.

 \begin{table}[H]
 \centering
\caption{Range for {\cmssm} parameters used in the initial brute-force generation of the training dataset.}
\label{tab:cmssm}
\centering
\begin{tabular}{c c c} 
 \hline \hline

 Parameter & Domain & Description \\ [0.5ex] 
 \hline
 $m_0$ & [0,10] TeV & Universal Scalar Mass \\ 
 $m_{1/2}$ & [0,10] TeV & Universal  Gaugino Mass \\
 $A_0$ & [-6$m_0$,6$m_0$] TeV & Universal Trilinear Coupling  \\
$\tan\beta$ & [1.5,50] & Ratio of Higgs VEVs \\
 $\text{sgn}(\mu)$ & 1 or -1 & Sign of Higgsino mass  \\ [1ex] 
 \hline \hline
\end{tabular}
\end{table}

 \begin{table}[H]
 \centering
         \caption{Range for {\pmssm} parameters used in the initial brute-force generation of the training dataset.. An additional pre-processing step was taken to manually remove the gaps in $|M_1|< 0.05$ TeV, $|M_2|< 0.1$ TeV, and $|\mu|< 0.1$ TeV before training.}
         \label{tab:pmssm}
         \centering
\begin{tabular}{c c c} 
 \hline \hline
 Parameter & Domain & Description \\ [0.5ex] 
 \hline
 $|M_1|$ & [0.05,4] TeV & Bino Mass \\ 
$|M_2|$ & [0.1,4] TeV & Wino Mass \\
$M_3$ & [0.4,4] TeV & Gluino Mass  \\
$|\mu|$ & [0.1,4] TeV & Bilinear Higgs Mass \\ 
$M_A$ & [0.1,4] TeV & Pseudo-scalar Higgs Mass \\
$|A_b|$ & [0,4] TeV & Trilinear Bottom Coupling  \\ 
$|A_t|$ & [0,4] TeV & Trilinear Top Coupling\\
$|A_\tau|$ & [0,4] TeV & Trilinear $\tau$ Coupling \\ 
$m_{\tilde{L}_1}$ & [0.1,4] TeV & 1st Generation L.h. Slepton Mass  \\
$m_{\tilde{L}_3}$ & [0.1,4] TeV & 3rd Generation L.h. Slepton Mass \\
$m_{\tilde{e}_1}$ & [0.1,4] TeV & 1st Generation R.h. Slepton Mass \\ 
$m_{\tilde{e}_3}$ & [0.1,4] TeV & 3rd Generation R.h. Slepton Mass   \\
 $m_{\tilde{Q}_1}$ & [0.4,4] TeV & 1st Generation L.h. Squark Mass  \\ 
 $m_{\tilde{Q}_3}$ & [0.2,4] TeV & 3rd Generation L.h. Squark Mass  \\ 
$m_{\tilde{u}_1}$ & [0.4,4] TeV & 1st Generation R.h. up-type Squark Mass  \\
$m_{\tilde{u}_3}$ & [0.2,4]  TeV & 3rd Generation R.h. up-type Squark Mass   \\
$m_{\tilde{d}_1}$ & [0.4,4] TeV & 1st Generation R.h. down-type Squark Mass   \\
$m_{\tilde{d}_3}$ & [0.2,4]  TeV & 3rd Generation R.h. down-type Squark Mass   \\
$\tan\beta$ & [1,60] & Ratio of Higgs VEVs\\ [1ex] 
 \hline \hline
\end{tabular}
\end{table}

\section{Approach}
Unlike other dimensional reduction methods, auto-encoders include a mapping back to the original space by design. We utilize a modified Sliced Wasserstein Autoencoder (SWAE) ~\cite{Kolouri:2018} which constrains the shape and consistency of the latent space to be suitable for efficient sampling, unlike standard non-generative  auto-encoders. The latent space created is abstract in the sense that there is no direct physical interpretation of its coordinates. But it is also structured such that points are encouraged towards a target distribution using a Sliced Wasserstein loss term. The network finds the most accurate way to encode this structure into reduced dimensionality by minimizing the reconstruction loss. New points that are generated in the low-dimensional latent space are then mapped back up to the original space, allowing for efficient sampling of the valid substructures in high dimensions. All feature columns are normalized to values between 0 and 1 before training.

\subsection{Why not a VAE?}
The Variational Auto-Encoder (VAE)~\cite{VAE} is a well-known generative framework using dimensional reduction, but complications in its structure make it non-ideal for our needs. Notably, the standard analytical form of the VAE's Kullback-Leibler (KL) loss assumes a Gaussian target distribution, which is achieved when the term is minimized at $\mu = 0$, $\sigma = 1$.  This pushes all points in the latent space to within 1 unit of the origin, which when combined with the variational nature of the VAE has been shown to cause information collapse during training: loss values can diverge or become degenerate because KL loss is not a true distance metric  ~\cite{Tolstikhin:2017,Howard:2021pos,zheng2018degeneration}. Additionally, because the KL loss term is defined only in terms of $\mu$ and $\sigma$, it is difficult to fit points to more complicated distributions such as those that are concentric or hollow. 

SWAEs come from a family of autoencoders specifically developed to avoid these issues ~\cite{Kolouri:2018, Howard:2021pos}, by replacing the KL loss term with the Sliced Wasserstein distance from optimal transport theory. The Sliced Wasserstein loss term fits latent points to individual points from the target distribution deterministically instead of variationally, mapping directly to latent coordinates instead of $\mu$ and $\sigma$. Thus SWAEs avoid the VAE's information collapse problem and can utilize complex target distribution shapes, making them a superior choice for our purposes. Wasserstein-style autoencoders have been shown in the past to produce higher quality samples in image generation tasks, compared to VAEs ~\cite{Kolouri:2018}.

\subsection{SAFESPAM} 
Using pytorch~\cite{NEURIPS2019_9015} and an ADAM optimizer~\cite{ADAM} \footnote{As well as code from the following packages: ~\cite{MLS,torchSWAE, 10.5555/1593511, 2020NumPy-Array, mckinney2010data, Waskom2021, hunter2007matplotlib}} , we train a modified SWAE to learn the mappings to and from the latent space, which we denote as $f$ and $g$ respectively . The mapping $\it{\{f: T \rightarrow L\}}$ from the theory space $\it{T}$ to the latent space $\it{L}$ is learned by the encoder, and the return mapping $\it{\{g: L \rightarrow T\}}$ is learned by the decoder. The full mapping of the SWAE can then be written:
\begin{align}
    T \xrightarrow{f} L \xrightarrow{g} T.
\end{align}

 In SAFESPAM, the primary modification is the addition of a clustering loss term which encourages valid points to cluster together. The loss function we use thus has three terms: the standard L2 reconstruction loss used by autoencoders to ensure accurate mapping, the Sliced Wasserstein loss term which fits latent points to a target distribution, and the clustering loss term, which pushes valid points toward the origin and invalid points away. Because the clustering loss term is aware of our experimental constraints, SAFESPAM is a supervised deep learning method, although weak-scale values do not enter the model as inputs. Our loss equation takes the following form:
\begin{equation}
\mathcal{L} = \alpha \mathcal{L}_\text{L2} + \beta \mathcal{L}_\text{Wasserstein} + \gamma \mathcal{L}_\text{Cluster}.
\end{equation}

\noindent
where $\alpha , \beta , \gamma $ are hyperparameters. Below we explain each term in more depth. \par

L2 loss, or MSE Mean loss, is calculated between the training data and the output of the SWAE, ensuring that the decoder accurately reconstructs the original inputs when mapping. It looks as follows: 
\begin{equation}
    \mathcal{L}_{L2} = \frac{1}{N}\sum_{i=1}^N(t_i - g(f(t_i)))^2,
\end{equation}

\noindent
where $N$ is the number of points per batch and $t$ the input theory point vector. \par

The Sliced Wasserstein loss is a formulation of the Wasserstein/earth-mover distance calculated along projected one-dimensional slices in latent space. It quantifies how the target distribution and latent distribution differ from one another.~\cite{Kolouri:2018, Howard:2021pos,torchSWAE} 
\begin{equation}
    \mathcal{L}_\text{Wasserstein} = \frac{1}{NM}\sum_{i=1}^N\sum_{j=1}^Mtc(\theta_i \cdot \tilde{s}_{k[j]},\theta_i\cdot f(t_{\ell[j]})).
\end{equation}
    Where $\theta_i$ represents the randomly sampled one-dimensional slices along which the marginal distributions are defined, $\tilde{s}_{k}$ and $f(t_{\ell})$ are random samples from the target distribution and the latent input data respectively, and $tc(\cdot,\cdot)$ represents the transport cost between the two distributions.

Our exploratory experiments indicated that for SAFESPAM, strong performance can often be achieved training only on valid points. In these instances, latent valid cores often take a convenient compact shape, useful for sampling and further structural exploration. Heuristically we find that utilizing concentric valid and invalid target distributions outperforms a bimodal normal target distribution where invalid points cluster around a point far from the latent origin. \par

Our clustering loss term takes inspiration from the Triplet Loss originally used in the fields of contrastive and metric learning ~\cite{Schroff:2015,NIPS2003_d3b1fb02}, with the goal of embedding a hierarchy of distances in the latent space, such that the distance between a desirable point and anchor point is minimized, and conversely the distance between an undesirable point and anchor point is maximized. For ease of sampling, we choose the latent origin as our anchor point. The mathematical expression per batch for our clustering is expressed as:
 
    \begin{equation}
        \mathcal{L}_\text{Cluster} = \frac{1}{N}\sum_{i=1}^N\left(a \delta_\text{valid} ||f(t_i)|| + \frac{b (1- \delta_\text{valid})}{||f(t_i)||}\right),
    \end{equation}
where $a,b$ are hyperparameters, $N$ is number of points per batch, and $t$ the input theory point vector. $\delta_\text{valid}$ is 1 for valid points and 0 for invalid points. This term encourages valid points to clusters near the latent origin and pushes invalids outward.  When training with only valid points, this term serves as an extra assurance that the valid portion of the latent space is sufficiently compact. When $\gamma$ = 0, a SAFESPAM network reverts to an unmodified SWAE. \par
\begin{figure}[H]
\centering
\includegraphics[scale=0.54]{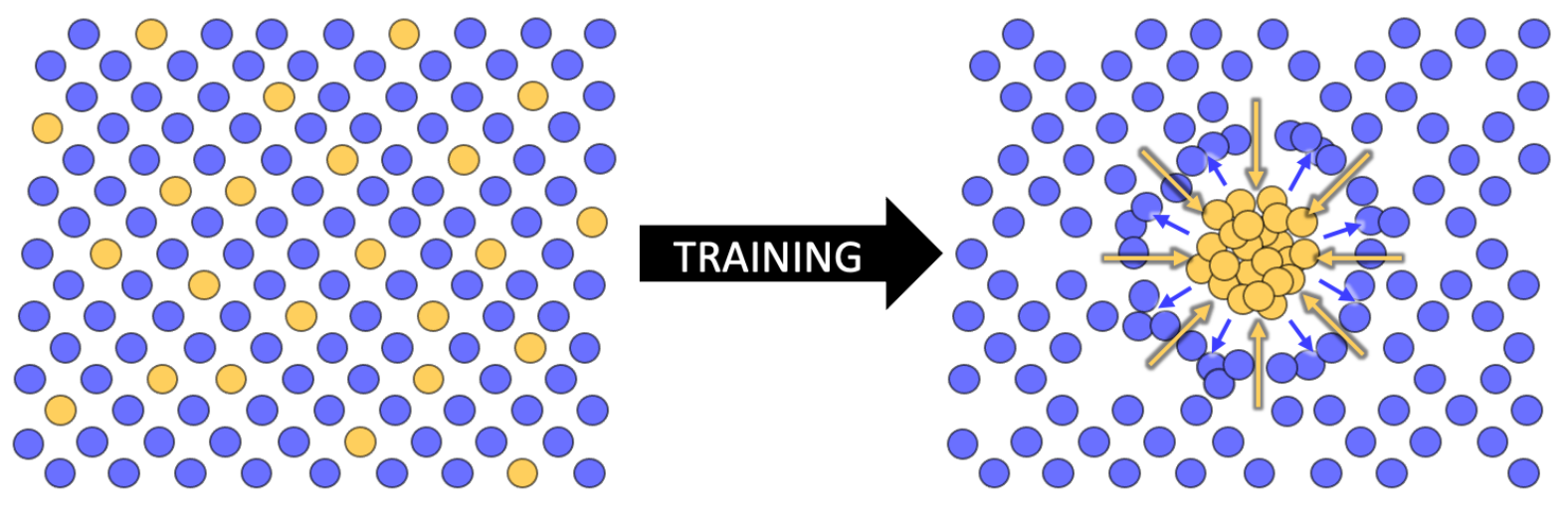}
\caption{A  cartoon illustration of the desired outcome of training on the latent space. Minimizing the clustering loss term pushes valid points (shown in orange) towards the latent origin while minimizing the Sliced Wasserstein loss fits those latent valid points to a target distributions. Shown in blue, invalid points (if present in training data) are pushed outward by the clustering loss term. The resulting clustering of valid points in the latent space allows for straightforward, high-efficiency sampling. Minimizing the L2 loss (not pictured) allows the network to successfully map back and forth between this structured latent space and the constrained MSSM parameter space.}
\label{fig:latent cartoon 1}
\end{figure}

Together, the Sliced Wasserstein and Clustering loss terms allow us to force favorable structure onto the latent space, controlling the distribution of valid and invalid points. Unlike standard non-generative auto-encoders, where poorly-defined latent spaces make sampling unfeasible, SAFESPAM enables efficient sampling by forcing desirable points to cluster together in one area; see Fig.~\ref{fig:latent cartoon 1} for a cartoon schematic.

\subsubsection{Sampling the latent space}
To sample, we generate points in the latent space and map them back to the original space using the network's decoder. Our method supports a variety of sampling options, including simple approaches such as uniformly sampling in the valid region.  We have found that Kernel Density Estimation (KDE) sampling is best at automatically capturing the shape of the latent valid core. KDE evaluates a kernel function over a dataset to estimate and sample from its probability density function. We utilized a top-hat kernel, with bandwidth $h$ such that $K(x;h)\propto 1$ if $x<h$, to estimate the density of our latent valid cores from known valid points, and sample the latent space accordingly. Using this kernel function, the density at a point in space $X$ in relation to a group of data points $x_j$ can be estimated as 
\begin{equation}
      \rho_K(X)=\sum^N_{j=1}K(X-x_j,h),
\end{equation}
 where $N$ is the number of points being used by the density estimation module. Essentially, this method approximates the density at each location by summing the number of data points that fall within the bandwidth distance of that point in space. KDE sampling works best in low-dimensional applications, making it well-suited to the latent space of our dimensional-reduction methods. 

The valid latent points can  be used to build a density estimate, via SciKit Learn's Density Estimation module~\cite{scikit-learn}, from which new points can be sampled. Because our mapping has already done the difficult task of clustering these points together in a reduced-dimensional representation,  KDE sampling in our latent space is more efficient than in the original space. 

\begin{figure}[H]
\centering
\includegraphics[scale=0.54]{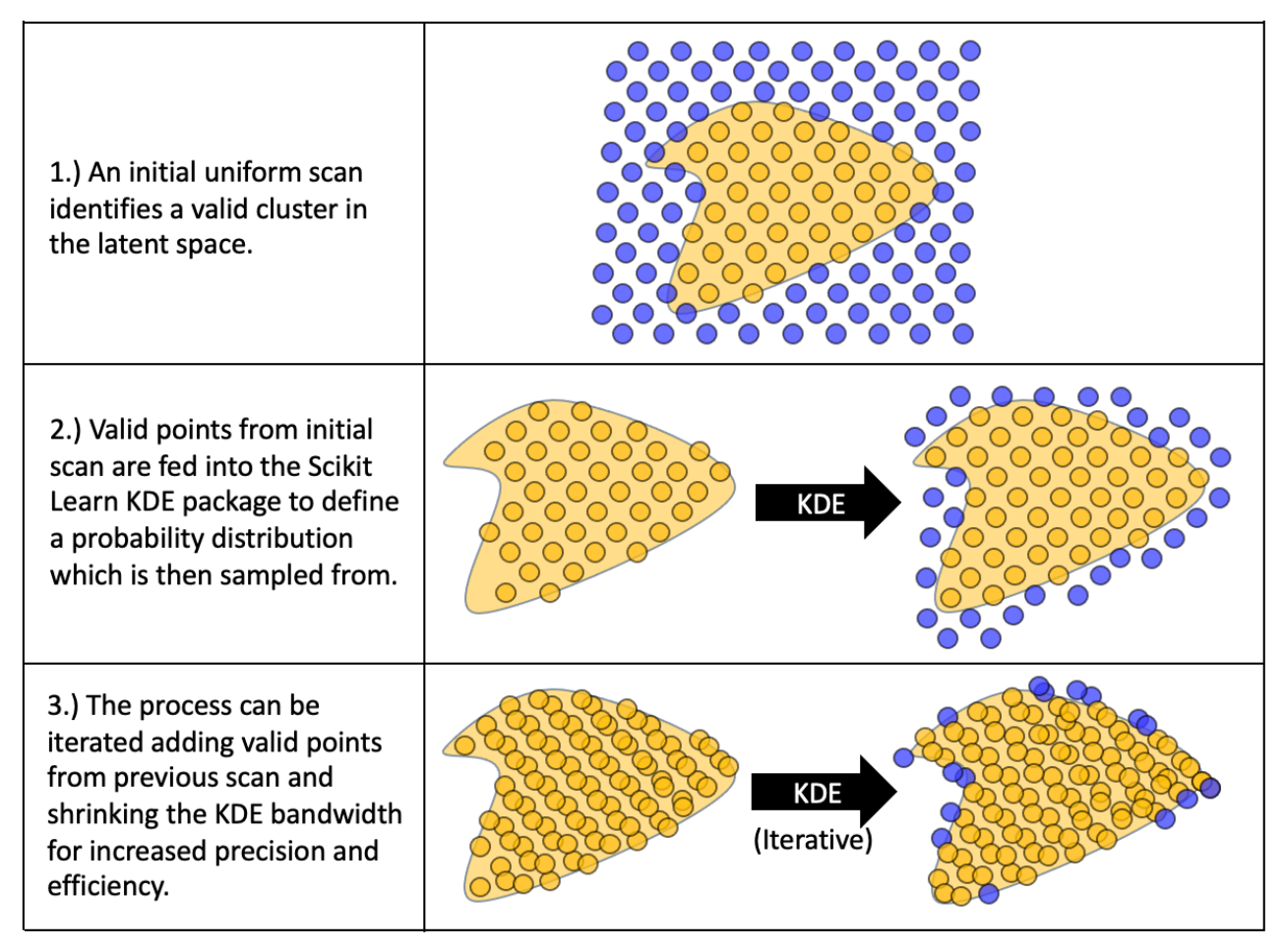}
\caption{A  cartoon illustration of the KDE sampling methodology. In CHUNC and CHUNC2, step 1 is omitted and known latent valids from the training data are used instead.}
\label{fig:KDE cartoon 1}
\end{figure}

\begin{figure}[H]
\centering
\includegraphics[scale=0.54]{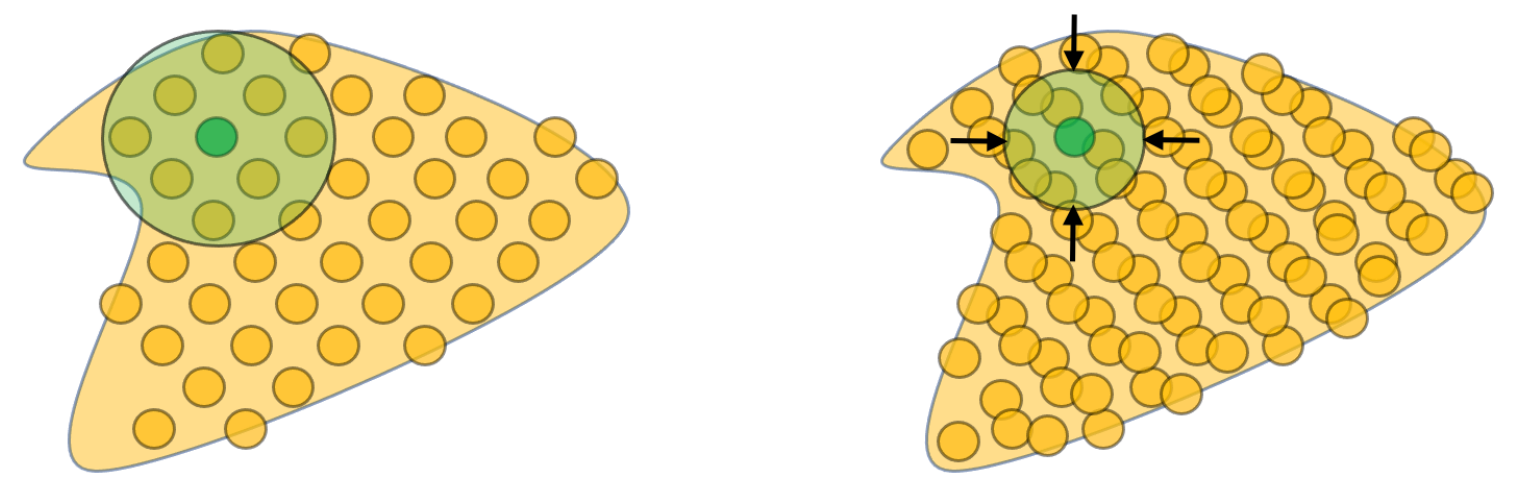}
\caption{A  cartoon illustration of the bandwidth shrinking procedure in our iterative KDE sampling methodology. As the number of known valid points grows with each successive iteration, we shrink the bandwidth used to estimate the density at each point in the latent space, allowing for higher precision.}
\label{fig:KDE cartoon 2}
\end{figure}

In some cases, the latent valid core takes an irregular shape that is more difficult to sample. To remedy this, we augment our basic KDE sampling approach iteratively. Valid points found in our first round of KDE sampling are added to the pool of points used to fit the next round of KDE sampling; see Fig.~\ref{fig:KDE cartoon 1}. As this iteration progresses, the bandwidth of the kernel is successively tightened, to sample finer structures more accurately; see Fig.~\ref{fig:KDE cartoon 2}. Iterative KDE serves as a means of transitioning from initial exploration of the latent space into more intensive sampling of the valid regions already found~\cite{Souza}.  This method can be repeated until a satisfactory efficiency is achieved.   \par 

While more advanced sampling options exist, such as training a normalizing flow to sample the potentially irregularly-shaped latent valid cores, we find KDE-style sampling to be much less labor-intensive and more robust at handling cases where there are disparities between the location of valid training data and valid sampled data. See, for example, the structures shown in Fig.~\ref{fig:enco/deco1}.

\par
While SAFESPAM does not guarantee unbiased sampling of the full valid subspace, it does offer significantly higher yields compared with naive sampling. In practice, this often means sampling with very high efficiency over an incomplete portion of the valid range. All SAFESPAM trials shown here used a sampling size of 10k points. \par 

\begin{figure}[H]
\centering
\includegraphics[scale=0.54]{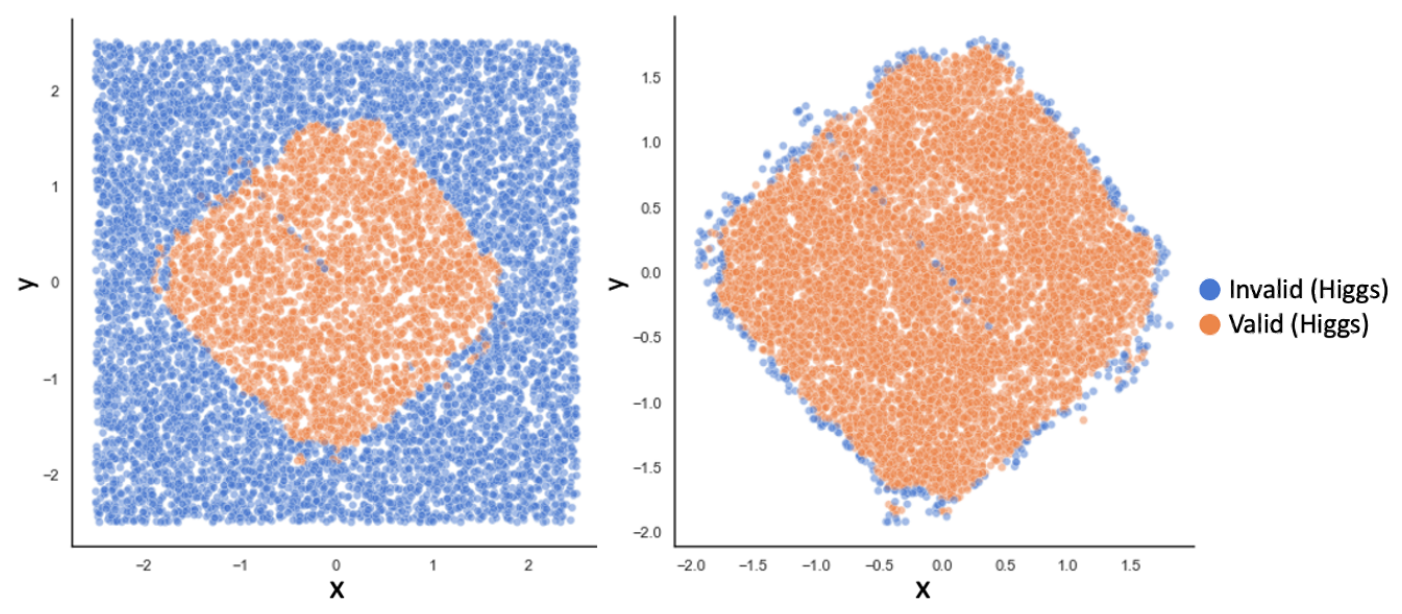}
\caption{An example of  KDE sampling in the {\pmssm}: An initial uniform scan of the latent space (left) demarcates the shape of the valid core, and the valid points found in that scan are utilized via kernel density estimation to sample the core efficiently (right).   White spaces represent areas that are avoided by KDE to improve efficiency.}
\label{fig:sampdemo}
\end{figure}

\subsection{CHUNC}
{\color{blue} Any function which reduces the dimensionality of its input has the potential to destroy information about the high-dimensional structures it seeks to describe. We anticipate such difficulties in high-dimensional spaces like the CMSSM and PMSSM where collapse to a two-dimensional latent space could cause significant portions of the valid sub-space to be overlooked with SAFESPAM-style methods. Thus, in an attempt to avoid this information loss, we introduce a similar methodology that operates without dimensional compression by imposing that the latent space is of the same dimension as the input space and that none of the layers in the network have a dimension smaller than the input.  

The motivation for this approach comes from the Data Processing Inequality (DPI) \cite{CoverThomas} which states that any function $f:T \rightarrow L$ acting on a space $T$ can only ever destroy or preserve information. This information is represented by the joint probability distribution $p(T,\theta)$, where $T$ is the theory space discussed in Section III and Section IV.2, and $\theta$ are the experimental constraints. The SWAE networks constructed here attempt to transform the unknown distribution $p(T,\theta)$ to one whose structure is known $p(L,\theta)$ while preserving the correlations between $L$ and $\theta$. The DPI states that the mutual information $I[T;\theta] \geq I[f(T);\theta] \geq I[g(f(T));\theta]$ with equality when no information loss has occured.  

To maximize $I[L;\theta] = I[f(T);\theta]$ then, we impose that the dimension of the space $L$ be the same as the dimension of $T$. } This comes at expense to the ease of sampling which is otherwise gained from utilizing a low dimensional latent space, and we also sacrifice the benefits of visualization. However our latent space still retains the desired quality that valid points cluster near the origin while invalid points are pushed outward. We demonstrate this method in {\cmssm} and {\pmssm} on a triple constraint:  Higgs, DM, and Neutralino LSP. We train on an equal mixture of valid and invalid points, which we fit to separate non-overlapping target distributions. Our Sliced Wasserstein loss term then becomes: 
\begin{equation}
    \mathcal{L}_\text{Wasserstein} = c \mathcal{L}^\text{Valid}_\text{Wasserstein} + d\mathcal{L}^\text{Invalid}_\text{Wasserstein}, 
\end{equation}
 where $c, d$ are hyperparameters. To reiterate, the role of the latent space is not to generate a low dimensional representation, but rather solely to transform the unknown distribution of the input space $p(T,\theta)$ to one $p(L,\theta)$ which is intentionally structured for sampling purposes, without any potential loss of information. We sample from this known distribution with KDE and map back to the original space, in the same manner as SAFESPAM. Because CHUNC retains usage of the Sliced Wasserstein loss term, the methods avoids the same VAE pitfalls as SAFESPAM. While the lack of dimensional reduction makes CHUNC superficially similar to normalizing flows, we believe our concentric contrastive target distributions are too complex to be utilized in a standard normalizing flow framework, which requires simple and well-understood base distributions ~\cite{Hollingsworth:2021sii}. 

\subsection{CHUNC2}
CHUNC2 is an alternative formulation of CHUNC which posits that further information loss might be avoided if the contrastive clustering is limited to a single additional categorical variable, $\theta$, allowing the rest of the latent space more freedom to preserve reconstruction capabilities. {\color{blue}Instead of pushing invalid points to infinity, the network learns to map to the joint space $L\times \theta$ so that the latent space is constrained as a simple Gaussian $p(L,\theta) = p(\theta)p(L|\theta) = p(\theta)p(\mu_L,\sigma_L|\theta)$ where the categorical variable $\theta$ interpolates between signal-like and background-like events.  As such, CHUNC2 replaces the clustering loss term mentioned in section 4.2 with a loss term built around an added categorical variable tasked with codifying whether a point is valid or invalid.} Thus our latent space is now $N+1$ dimensional, where the new dimension serves as a predictor of a point's validity: \begin{itemize}
 \item \textbf{Binary $L^2$ Loss} - The latent categorical variable $z_i$, is constrained through an $L^2$ loss to match the category $\theta_i = \{s,b\}$ of each event,
   \begin{equation}
       \mathcal{L}_{\mathrm{bin}}(z,\theta) = \frac{1}{N}\sum_{i=1}^N(z_i - \theta_i)^2.
   \end{equation}
    \end{itemize}
 In the following studies, the categorical variable is trained to map to 1.0 for valid points and 0.0 for invalid points. {\color{blue}It is of course straightforward to generalize this approach to one which $\theta$ represents a multi-class problem, but we do not pursue this here since our task is a binary one (i.e. with our triple-constraint task we are not interested in generating points that only meet some but not all of the individual constraints). } Sampling utilizes KDE on latent points which fall into the most valid categorical bin to generate new latent points in 6D or 20D respectively, which are mapped back to the original space. Since CHUNC2's latent space is of a higher dimensionality than the parameter space, normalizing flows would not be able to emulate this methodology. 
 
\section{Results}
\subsection{SAFESPAM}
We trained a series of Modified SWAEs on {\cmssm} and {\pmssm} data using Higgs and DM constraints separately. Unless otherwise noted, all SAFESPAM models were trained on valid-only data\footnote{Note that using valid-only data is not a requirement of the method.}, and latent points were constrained to a truncated normal target distribution by minimizing the Sliced Wasserstein loss. The latent space was sampled with an initial uniform scan whose location was determined by the latent position of valid training data. Valid points from this scan were then used to define a probability density sampled with single-round KDE sampling, using a tophat kernel for increased precision. Iterative KDE was utilized to improve results in cases with the Dark Matter experimental constraint. Below, we present the results for Higgs mass constraints in the {\cmssm} and {\pmssm} cases, followed by dark matter relic density constraints in each space, and an analysis of the ability of our method to simultaneously satisfy both constraints. We present visuals of the sampled points in the latent space to confirm the presence of a central valid core. We also include histograms comparing our generated valid data to our valid training data to qualify to what extent our sampling has introduced unwanted bias or excluded viable regions of the space.  We then report sampling efficiencies, defined as the fraction of points from the sample which adhere to the desired constraint. Relevant hyperparameters are listed in Appendix A.  All SAFESPAM model-training for this project took place on a 2017 macbook pro without use of GPU. 
\par 
{\color{blue} We introduce two additional metrics to better quantify the successes and shortcomings of this method, described as follows:
\newpage
    \begin{itemize}
    \item Core-forming Metric
    \begin{itemize}
    \item This metric measures how successfully the generated valid points match the core shape suggested by the position of latent valid training data. We calculate the Sliced Wasserstein distance between randomly selected points from the latent distribution of valid training points and the latent distribution of valid generated points. Latent points are normalized before calculation to account for differing target distribution sizes between trials. A low value suggests that the generated valid points are occurring where we expect them to be, whereas a higher value suggests potential disparities between the latent distributions of training valids and generated valids. Multiple calculations of this value are averaged out to account for the random nature of the slices. All SAFESPAM metrics were calculated by taking the average over 10 iterations, using 200 projections. It takes the form: 
    \begin{equation}
    \mathcal{L}_\text{Wasserstein} = \frac{1}{NM}\sum_{i=1}^N\sum_{j=1}^Mtc(\theta_i \cdot \tilde{gl}_{k[j]},\theta_i\cdot f(tl_{\ell[j]})),
    \end{equation}
    Where $\theta_i$ represents the randomly sampled one-dimensional slices along which the marginal distributions are defined, $\tilde{gl}_{k}$ are the normalized valid latent points generated by the model and $f(tl_{\ell})$ are random normalized latent representations of samples from the valid training dataset, and $tc(\cdot,\cdot)$ represents the transport cost between the two distributions.
    \end{itemize}
    \item Incompleteness Metric
    \begin{itemize}
    \item This metric measures how successfully the generated valid points account for the full distribution of expected valid points seen in the training data. We calculate the Sliced Wasserstein distance between randomly selected points from the GUT distribution of valid training points and the GUT distribution of valid generated points (excluding valid generated points outside the original range the training data was generated with). All points are normalized before calculation to account for the different ranges used in data generation in the {\cmssm} and {\pmssm}.  A low value indicates we are sampling without much bias, whereas a higher value suggests we are introducing more significant bias when sampling. Multiple calculations of this value are averaged out to account for the random nature of the slices.  All SAFESPAM metrics were calculated by taking the average over 10 iterations, using 200 projections. It takes the form: 
    \begin{equation}
    \mathcal{L}_\text{Wasserstein} = \frac{1}{NM}\sum_{i=1}^N\sum_{j=1}^Mtc(\theta_i \cdot \tilde{gg}_{k[j]},\theta_i\cdot f(tg_{\ell[j]})),
    \end{equation}
    Where $\theta_i$ represents the randomly sampled one-dimensional slices along which the marginal distributions are defined, $\tilde{gg}_{k}$ are the normalized valid GUT points generated by the model and $f(tg_{\ell})$ are random normalized samples from the valid training dataset, and $tc(\cdot,\cdot)$ represents the transport cost between the two distributions.
    \end{itemize}
    \end{itemize}}

\subsubsection{Higgs cases}

\begin{figure}[H]
\centering
\includegraphics[scale=0.35,center]{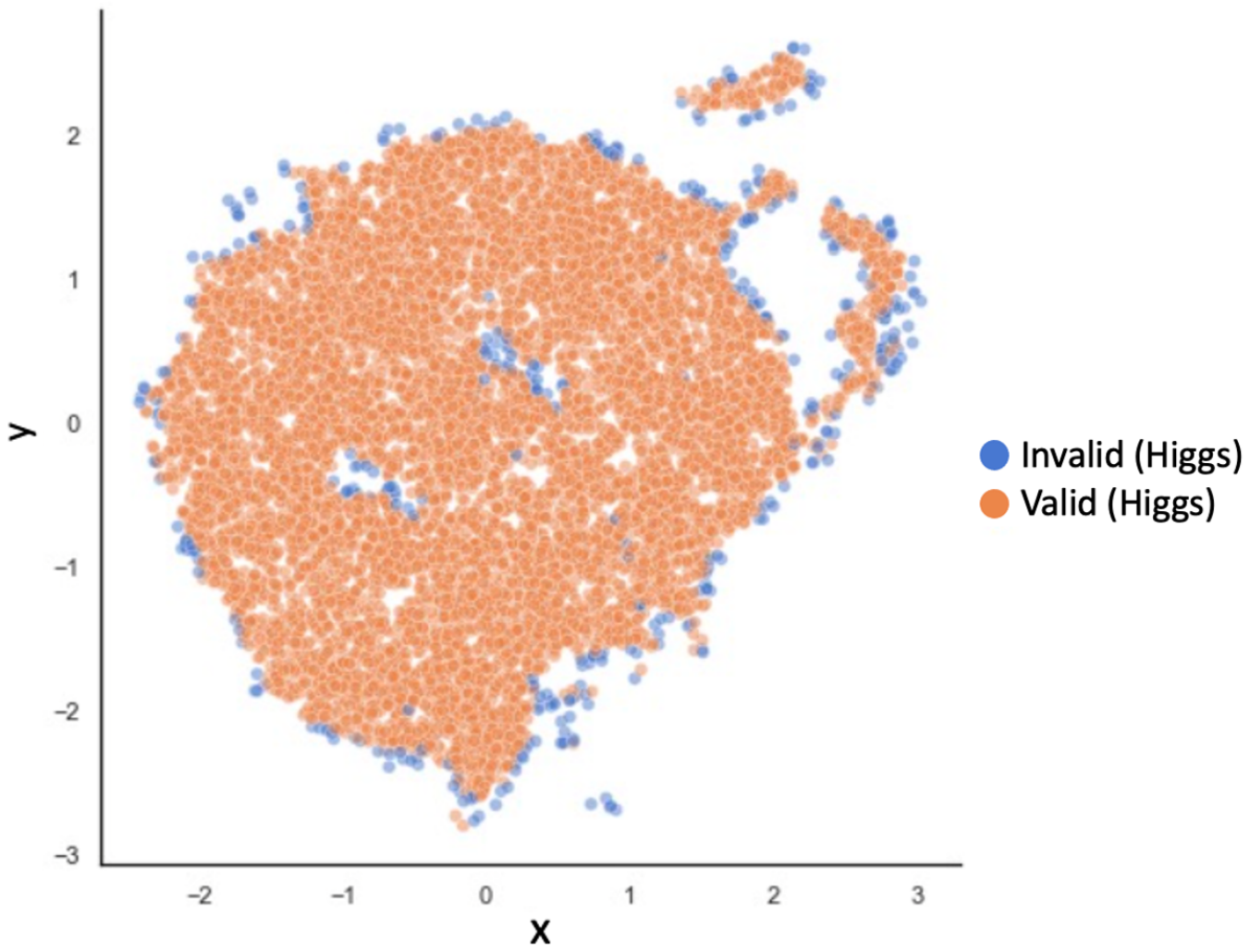}
\caption{Distribution of sampled points in the two-dimensional latent space, for the case of learning to generate {\cmssm} points which satisfy the Higgs mass constraint.   Points which satisfy (fail) the Higgs mass constraint are marked Valid (Invalid). Invalid regions in the center are partially avoided due to KDE sampling.}
\label{fig:Higgslatcmssm}
\end{figure}

\begin{figure}[H]
\centering
\includegraphics[scale=0.54,center]{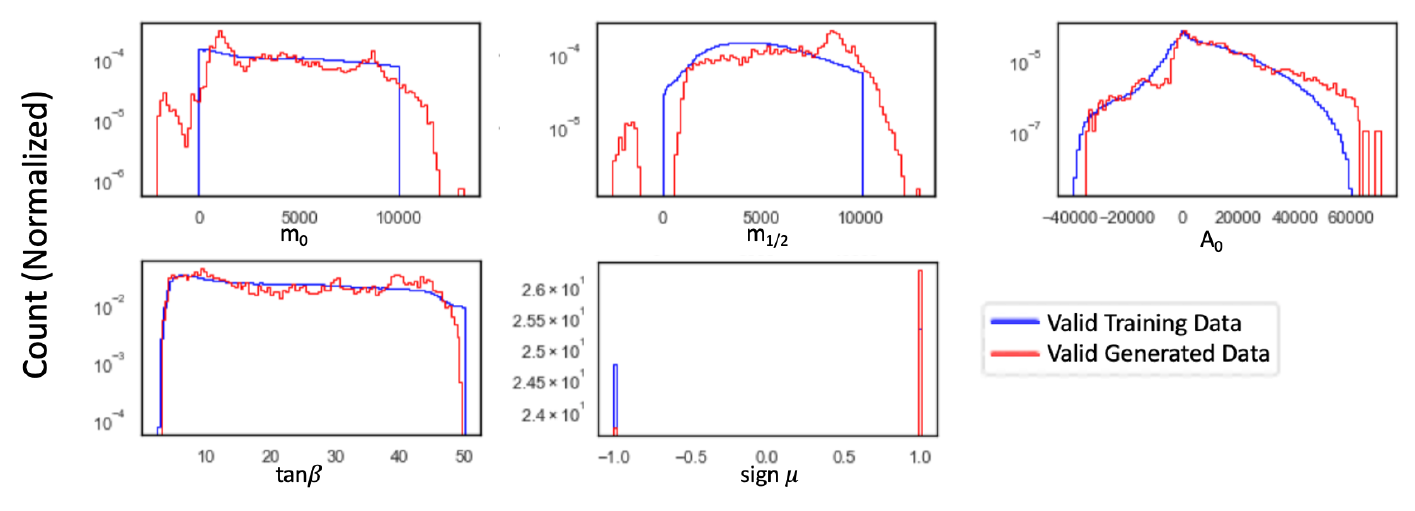}
\caption{Distribution of SAFESPAM-generated points (red) in the {\cmssm} in each of the five {\cmssm} parameters, under the Higgs mass constraint, compared to the distribution of the training dataset (blue).}
\label{fig:Higgs cmssm}
\end{figure}

 Our method had little trouble creating a valid core using Higgs mass as the experimental constraint in the {\cmssm}; see Fig.~\ref{fig:Higgslatcmssm}, which allowed for highly efficient sampling efficiency (0.957) compared to a naive scan (0.347).  Small incursions of invalid points in the core are largely avoided by KDE sampling.  The distribution of SAFESPAM-sampled points in the space of {\cmssm} parameters is shown in Fig.~\ref{fig:Higgs cmssm}, where the newly generated points reasonably match the distribution in the original training set.

 In the case of the \pmssm, SAFESPAM was again able to structure the latent space for efficient sampling. We note that the specific form of the latent space demonstrates significant sensitivity to the hyperparameters such as the shape of the target distribution, length of training, and inclusion or exclusion of valid points, though without a significant change in efficiency.

\begin{figure}[H]
\centering
\includegraphics[scale=0.6,center]{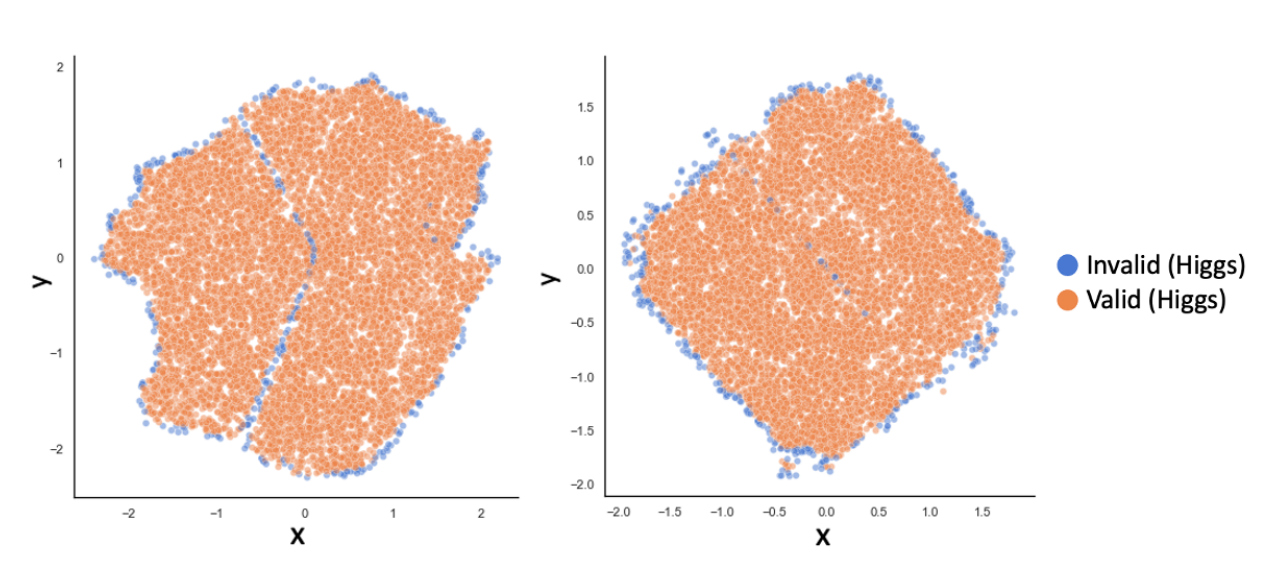}
\caption{ Distribution of sampled points in the two-dimensional latent space, for the case of learning to generate {\pmssm} points which satisfy the Higgs mass constraint.  
Points which satisfy (fail) the Higgs mass constraint are marked Valid (Invalid). Shown are points sampled via KDE, where invalid regions in the center are partially avoided due to KDE sampling. Trial 1 is shown on the left and Trial 2 on the right, showcasing the sensitivity of the latent space structure to the hyperparameters.}
\label{fig:pmssmlatHiggs}
\end{figure}

\begin{figure}[H]
\centering
\includegraphics[scale=0.54,center]{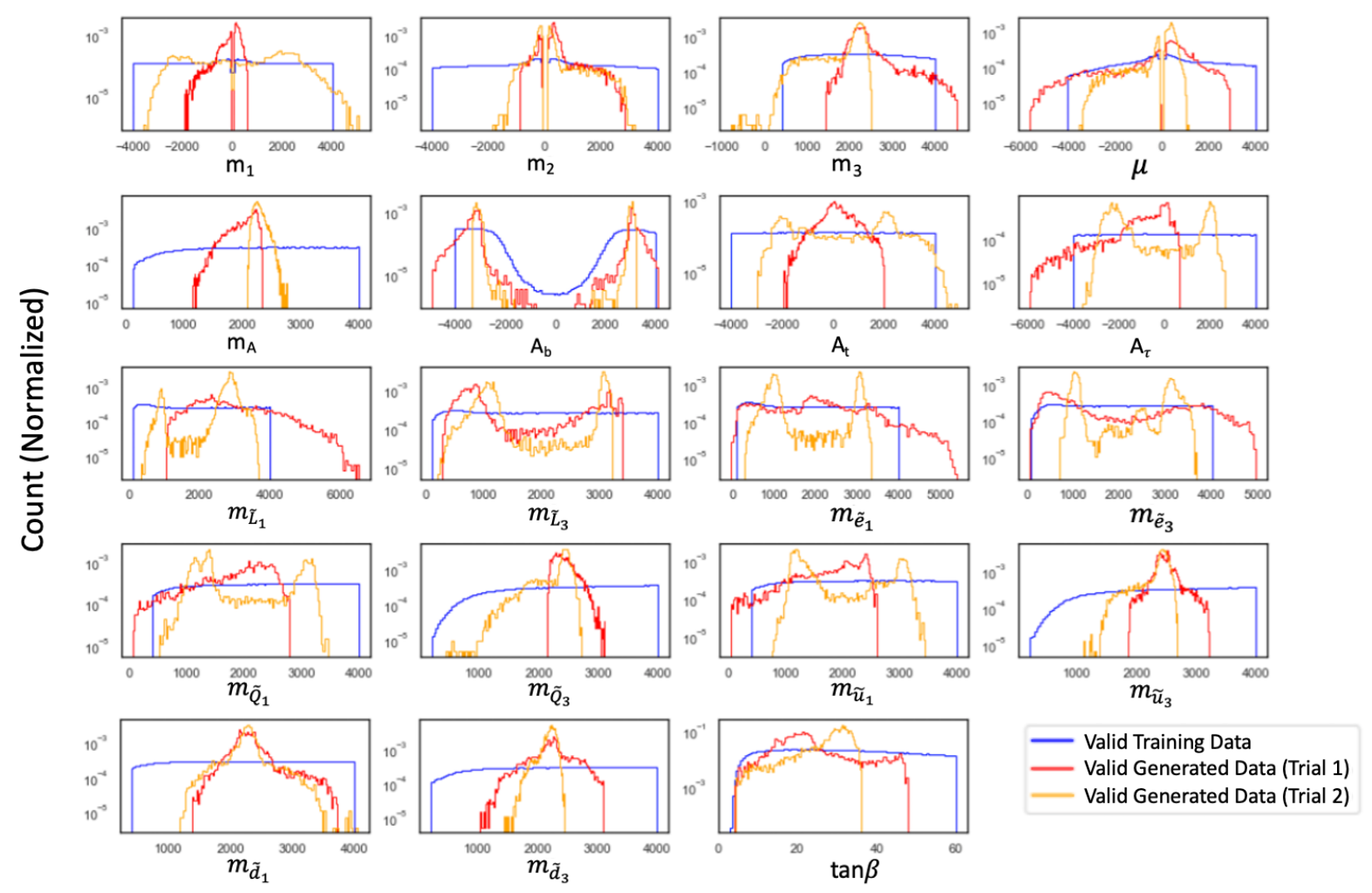}
\caption{Distribution of SAFESPAM-generated points (red for Trial 1, orange for Trial 2) in the {\pmssm}  in each of the nineteen {\pmssm} parameters, compared to the distribution of the training dataset (blue). While both trials achieve strong efficiency, they sample different portions of the parameter space.}
\label{fig:pmssm Higgs}
\end{figure}

 As a demonstration, we present results for two separate networks trained for the {\pmssm} Higgs constraint, labeled Trials 1 and 2 respectively. Trial 1 was trained first on a combination of valid and invalid points fit to a concentric distribution (a normal distribution with an ablated annulus to create two distinct regions), before being trained further with only valid points fit to a truncated normal. Trial 2 was trained only on valid points fit to a truncated normal, but trained for four times as many epochs. Figure~\ref{fig:pmssmlatHiggs} shows the distribution of the valid points in the latent space for each trial, with clear variation in the structure of the latent space.  KDE sampling of the core achieves an efficiency of 0.965 (0.961) for Trial 1 (Trial 2), compared to naive sampling efficiency of 0.171.  Distribution of the generated points in the {\cmssm} parameter space for both trials is shown in Fig.~\ref{fig:pmssm Higgs}.  Although neither trial generated the full range of valid points seen in the training data, future methods might make use of ensemble sampling to remedy this. 

\subsubsection{Dark Matter Cases}
\begin{figure}[H]
\centering
\includegraphics[scale=0.65,center]{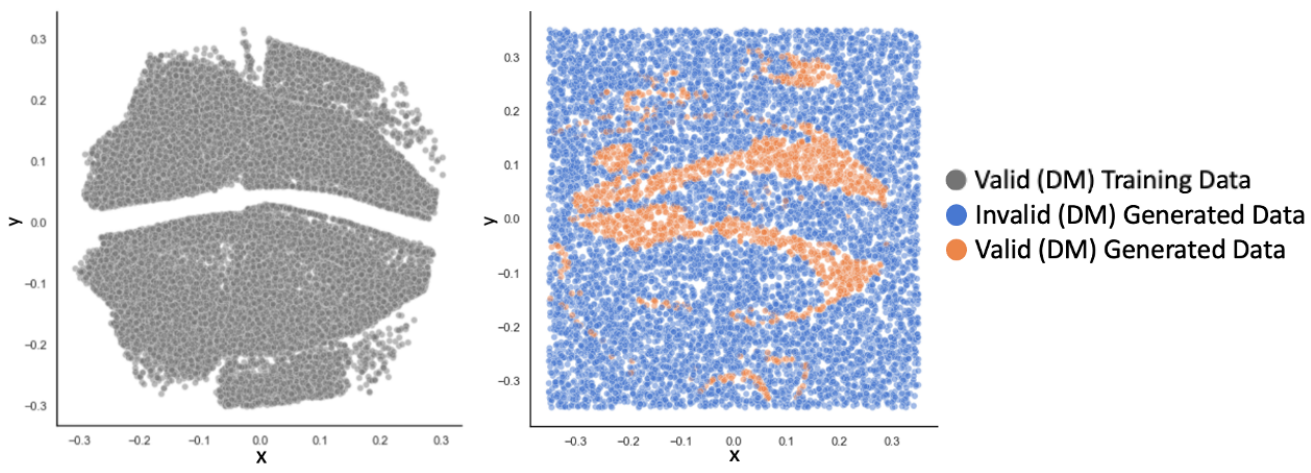}
\caption{Distributions in the two-dimensional latent space for the {\cmssm} dark matter study, where points which satisfy (fail) the dark matter constraint are marked Valid (Invalid). Left shows the distribution of valid points in the training data, with a well-formed core.  However, many new points sampled from this core do not satisfy the dark matter constraint (right), perhaps due to insufficient resolution of DM-valid features in the training data. Successive KDE iterations allowed for precise, efficient sampling in the remaining regions of high validity (see Figs~\ref{fig:DM cmssm lat} and~\ref{fig:DMpmssm1}).}
\label{fig:enco/deco1}
\end{figure}

\begin{figure}[H]
\centering
\includegraphics[scale=0.54,center]{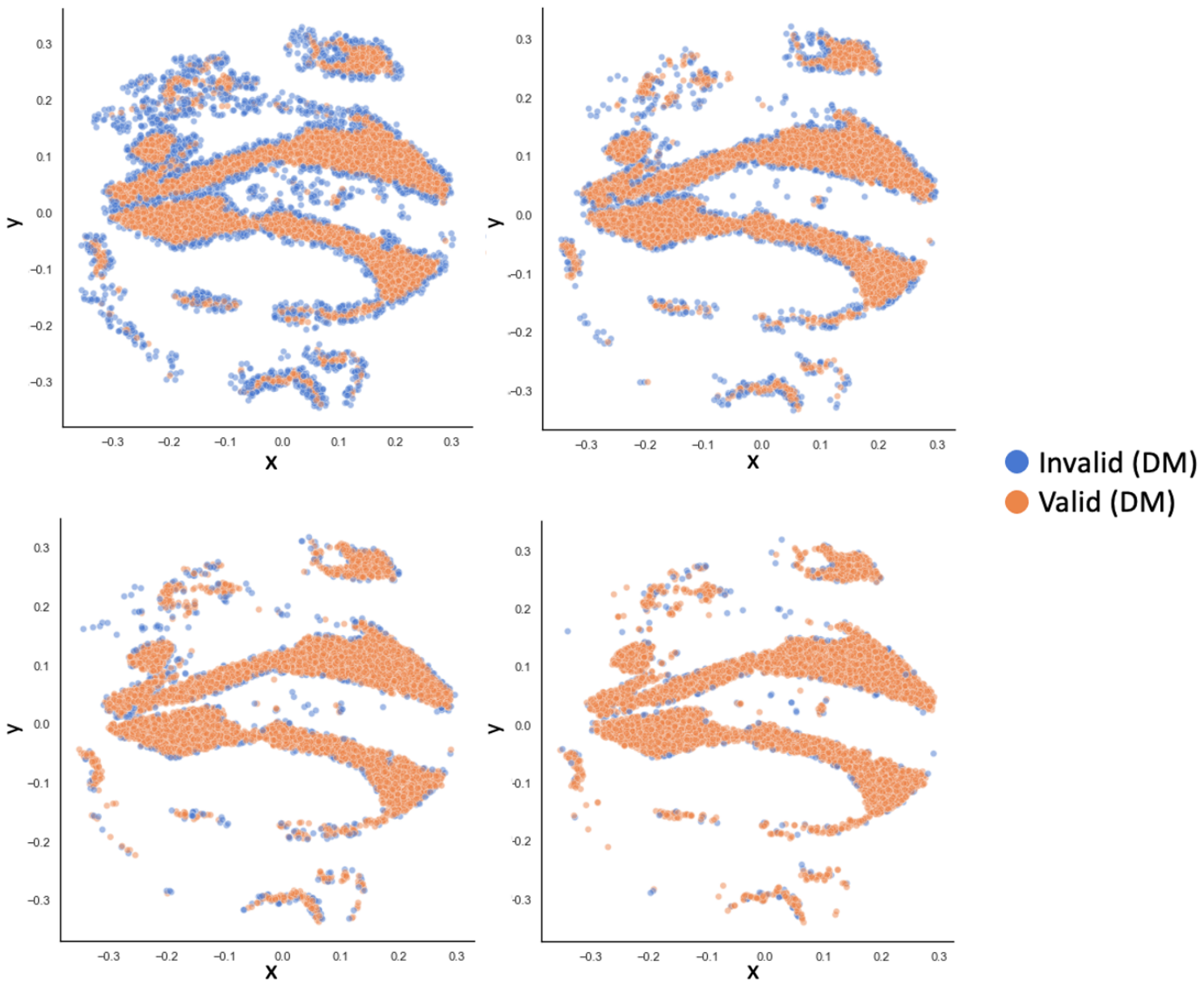}
\caption{Distributions of sampled points in the two-dimensional latent space in four successive rounds of sampling for the {\cmssm} dark matter case. Points which satisfy (fail) the dark matter constraint are marked Valid (Invalid). In each round, newly discovered valid points are used to update the kernel density estimation of valid regions of the latent space, which is used to perform the next round of sampling.  As a result, the `crust' of invalid points shown in blue shrinks away with each successive iteration. The efficiencies for these 4 rounds are 0.724, 0.887, 0.945, and 0.972 respectively.}
\label{fig:DM cmssm lat}
\end{figure}

\begin{figure}[H]
\centering
\includegraphics[scale=0.6,center]{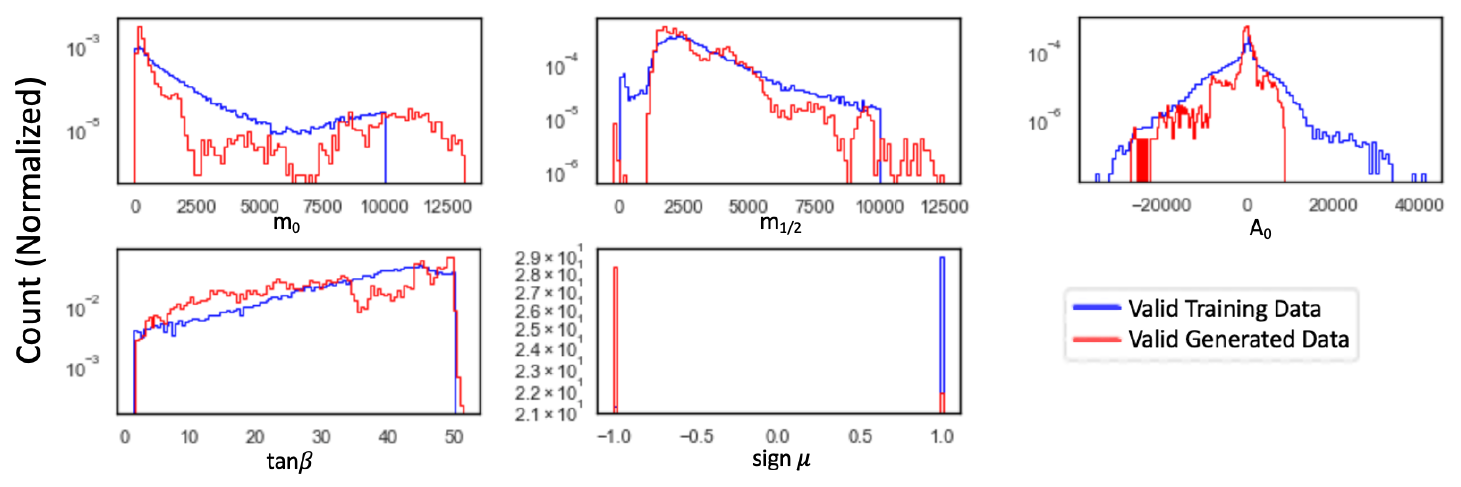}
\caption{Distribution of SAFESPAM-generated points in the {\cmssm} (red) in each of the five {\cmssm} parameters under the dark matter constraint, compared to the distribution of the training dataset (blue).}
\label{fig:DM cmssm hist}
\end{figure}

\begin{figure}[H]
\centering
\includegraphics[scale=0.50,center]{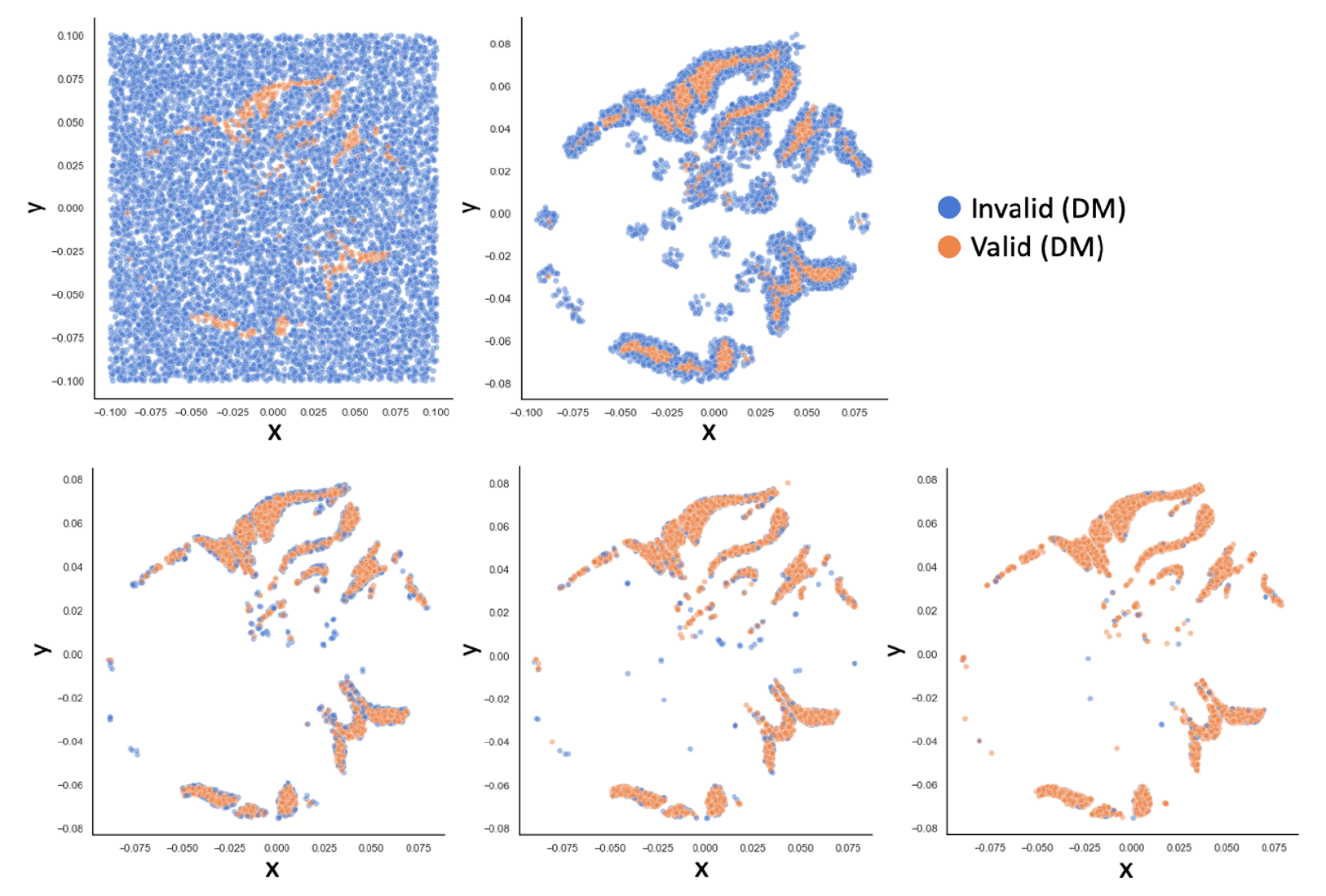}
\caption{Distributions in the two-dimensional latent space for the {\pmssm} dark matter study, where points which satisfy (fail) the dark matter constraint are marked Valid (Invalid). Shown are distributions of sampled points in successive rounds, where in each round valid points from previous rounds are used to perform a kernel density estimation of valid regions of the latent space for the next round of sampling. The initial scan in the top left frame samples via uniform distribution over a region determined by the position of latent valid training data. The efficiency for each round is 0.049, 0.432, 0.881, 0.950, 0.979, respectively.}
\label{fig:DMpmssm1}
\end{figure}

\begin{figure}[H]
\centering
\includegraphics[scale=0.6,center]{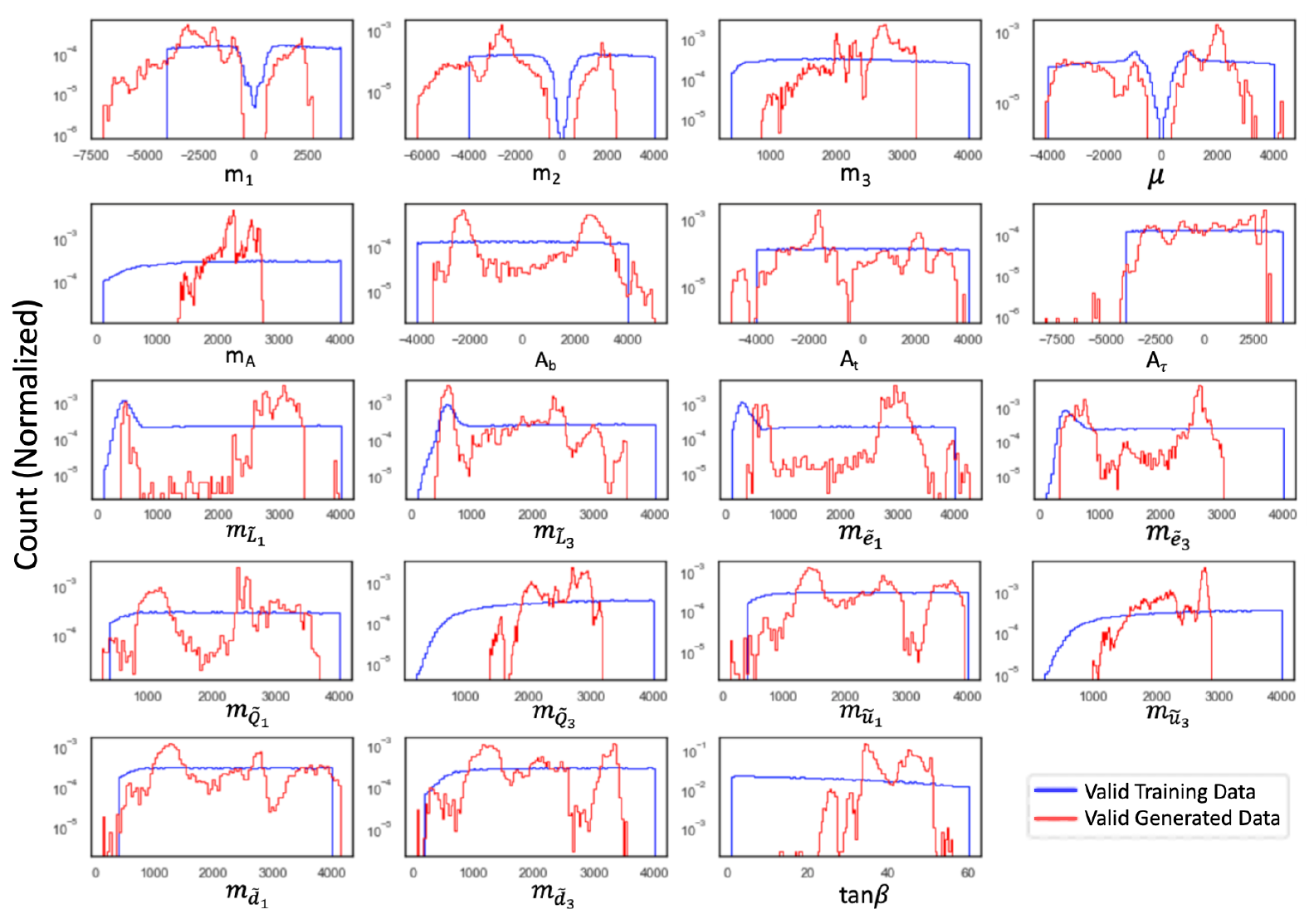}
\caption{Distribution of SAFESPAM-generated points in the {\pmssm} (red) in each of the nineteen {\pmssm} parameters under the dark matter constraint, compared to the distribution of the training dataset (blue).}
\label{fig:DM pmssm 1}
\end{figure}

 Our dark matter trials had significantly more difficulty creating well-structured latent spaces. While the method was successful in clustering together valid training data, sampling in this vicinity resulted in many points which do not satisfy the constraints; see Fig.~\ref{fig:enco/deco1}.  However, a subset of points from the initial core allowed for efficient generation of new valid points, as discovered after iterating the KDE sampling procedure four times.   This led to high efficiency of 0.972(0.979), compared to naive sampling efficiency of 0.05534(0.006506) in the {\cmssm} (\pmssm). See Figs~\ref{fig:DM cmssm lat} and~\ref{fig:DMpmssm1}. The iterative nature of our sampling method allowed us to  probe these valid cores efficiently despite their irregular shape. As in the Higgs trials, this method had more success in preserving the valid range of points from the training data when used in the {\cmssm} (Fig. \ref{fig:DM cmssm hist}), and less so in the {\pmssm} (Fig. \ref{fig:DM pmssm 1}). We speculate that the difficulty here may result from the particularly complex and narrow structures in the {\cmssm} and {\pmssm} created by the extreme narrowness of the dark matter constraint as applied to these spaces: we believe our valid-only training dataset lacked the sufficient resolution required to encode their boundaries accurately in the latent space.

\subsubsection{Adding secondary constraints without additional training}
\begin{figure}[H]
\centering
\includegraphics[scale=0.55,center]{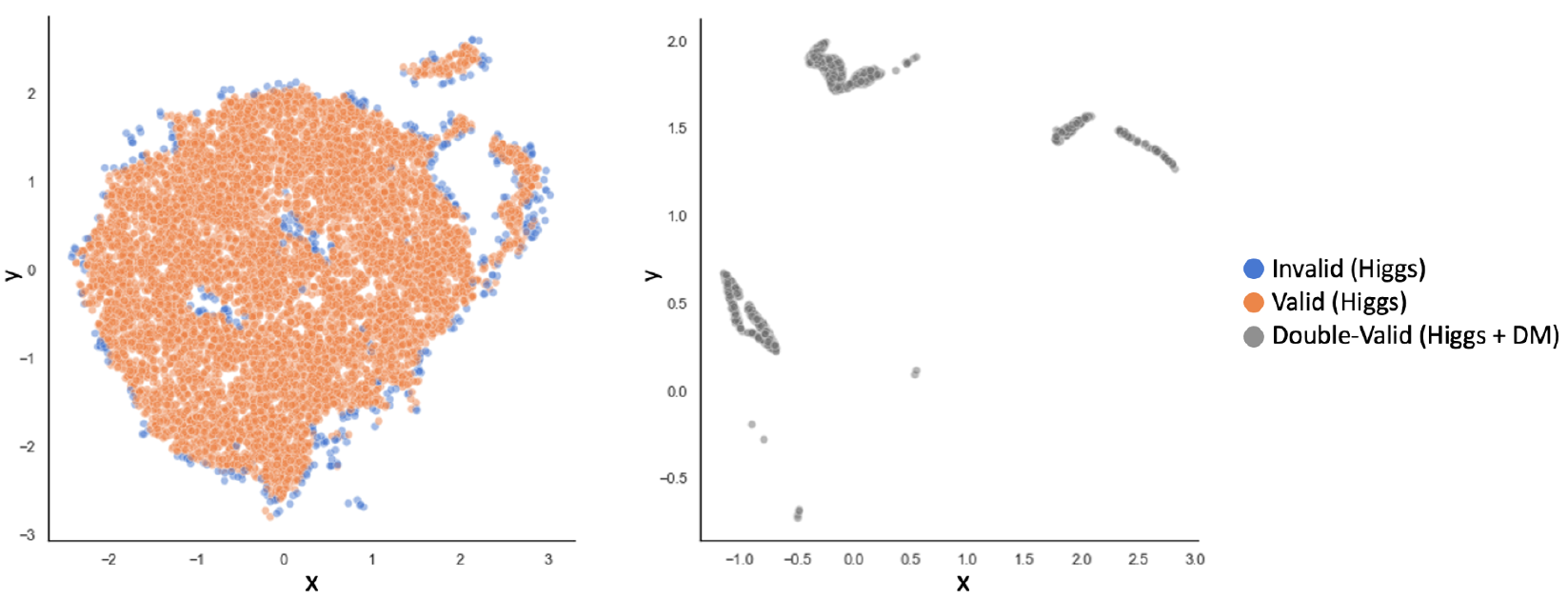}
\caption{ Distributions in the two-dimensional latent space for the {\cmssm} study, where points which satisfy (fail) the Higgs mass constraint are marked Valid (Invalid), and valid points which also satisfy the dark matter constraint are marked as Double-Valid. In the left, the initial scan identifies a core of valid Higgs points. On the right are the newly generated points which satisfy both constraints.}
\label{fig:cmssm Double}
\end{figure}

\begin{figure}[H]
\centering
\includegraphics[scale=0.70,center]{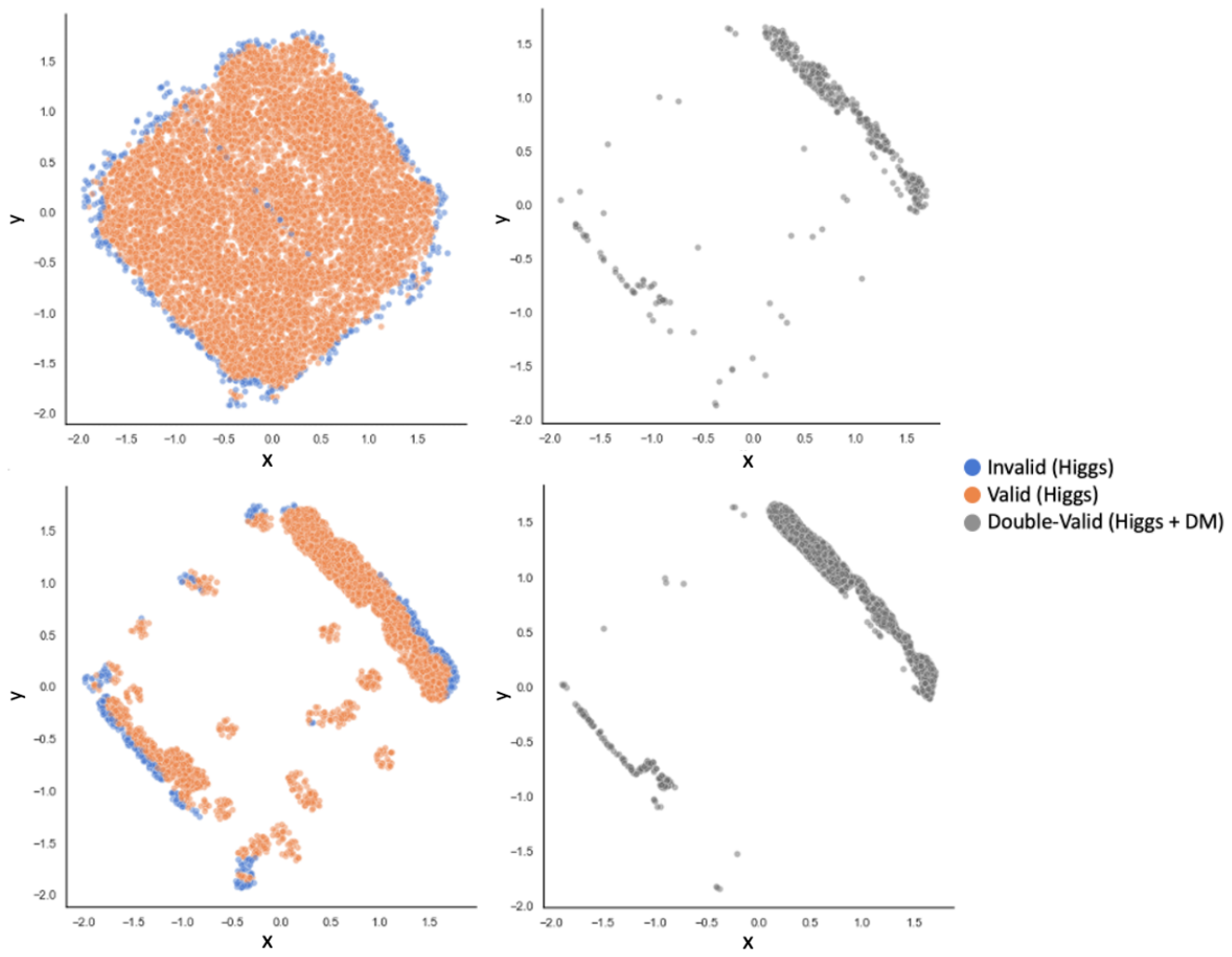}
\caption{ Distributions in the two-dimensional latent space for the {\pmssm} study, where points which satisfy (fail) the Higgs mass constraint are marked Valid (Invalid), and valid points which also satisfy the dark matter constraint are marked as Double-Valid. In the top left, the initial scan identifies a core of valid Higgs points. In the top right, the subset of those points which also satisfy the dark matter constraints are identified. In the bottom left are points generated using a KDE built from the double-valid points, and in the bottom right are the new points which satisfy both constraints.}
\label{fig:pmssm Double}
\end{figure}

The structure of the SAFESPAM latent space allows us to apply additional constraints {\it in media res} when sampling. We identify latent points which meet these additional constraints via initial scan or previous sampling and use KDE sampling to draw new points from their vicinity in the latent space, without providing any additional information to the auto-encoder about which points satisfy these secondary constraints.  We demonstrate this in the {\cmssm} (Fig. \ref{fig:cmssm Double}) and {\pmssm} (Fig. \ref{fig:pmssm Double}) by re-sampling our Higgs models for points that have both valid Higgs and DM values, i.e. a joint constraint.  Although our method does not guarantee that such `double-valid' points will exist in the latent space or form any neat self-contained substructure, we found our KDE sampling was sufficiently robust to sample the clusters of points that were present in the latent space, achieving a significantly increased efficiency compared to the naive scan, and doing so without requiring any additional training. This property gives our maps an added versatility that many non-mapping generative models and efficient search methods are unequipped for. For example, one can add additional experimental constraints or look for theory points from a specific dark matter annihilation class just by selectively sampling regions of the map where those characteristics are known to exist.   Our method's dimensional reduction capabilities also give it a significant advantage in ease of sampling these added constraints compared to other mapping-based generative methods such as normalizing flows which could in theory attempt something similar. We see significant increases in efficiency with single-round KDE up to 0.3191 (0.5621) in the {\cmssm} ({\pmssm}) compared to a naive value of 0.004467 (0.01892). Additional iterations of KDE may improve these results further. 

Table~\ref{table:eff} summarizes the generation efficiency of experimentally valid points using SAFESPAM and naive scanning.

\begin{table}[H]
\centering
\caption{ Comparison of the efficiency of generation of experimentally-valid points in naive scanning versus SAFESPAM, for several experimental constraints (Higgs mass $m_H$, dark matter relic density $\Omega_{DM}$, or both) and two theoretical spaces (5-dim {\cmssm}, 19-dim {\pmssm}).}
\label{table:eff}
\begin{tabular}{c |c| c c l} 
 \hline \hline
 & &\multicolumn{2}{c}{Method}\\
  Space & Constraint & Naive scan &  SAFESPAM  & Notes \\ [0.5ex] 
 \hline
 {\cmssm} & $m_H$ & 0.347 & 0.957 \\ 
 {\pmssm} & $m_H$ &  0.171  & 0.965 & Trial 1  \\  
   &  &    & 0.961 & Trial 2 \\  

 \hline
 {\cmssm} & $\Omega_{DM}$ & 0.0065  & 0.972 & Iterative\\
 {\pmssm} & $\Omega_{DM}$ & 0.0553  & 0.979 & Iterative \\
\hline
 {\cmssm} & $m_H \cap \Omega_{DM}$ & 0.0045 & 0.319 & Trained for $m_H$  \\ 
 {\pmssm} & $m_H \cap \Omega_{DM}$ & 0.0189  & 0.562 & Trained for $m_H$ \\ 
\hline \hline
\end{tabular}
\end{table}

\begin{table}[H]
\centering
\caption{{\color{blue} Comparison of the Core-Forming and Incompleteness Metrics for various SAFESPAM trials for several experimental constraints (Higgs mass $m_H$, dark matter relic density $\Omega_{DM}$) and two theoretical spaces (5-dim {\cmssm}, 19-dim {\pmssm}).}}
\label{table:eff2}
{\color{blue}\begin{tabular}{c |c| c c c l} 
 \hline \hline

  Space & Constraint & Core-Forming &  Incompleteness & Notes \\ [0.5ex] 
 \hline
 {\cmssm} & $m_H$ & 0.0037 & 0.0043  \\ 
 {\pmssm} & $m_H$ &  0.0167 & 0.0202 & Trial 1  \\  
   &  &  0.0081  & 0.0135 &  Trial 2 \\  
 \hline
 {\cmssm} & $\Omega_{DM}$ & 0.0051  & 0.0152 &  Iterative\\
 {\pmssm} & $\Omega_{DM}$ & 0.0403 & 0.0355 &  Iterative \\
\hline
\end{tabular}}
\end{table}

\subsection{CHUNC and CHUNC2 for Triple Constraint}
The {\cmssm} and {\pmssm} training datasets for CHUNC and CHUNC2 each consist of approximately 10K points, evenly split between invalid points and triple-valid points, which satisfy the Higgs mass, dark matter relic density and LSP constraints defined in section 2.

The higher-dimensional latent space requires a slightly different KDE sampling method. Rather than performing a uniform scan to identify a potential core, we use the position of known latent valid points from the training set to perform the kernel density estimation. CHUNC attempts to cluster all invalid points on an $N$-dimensional spherical shell of radius 1 with of thickness of 0.6, and valid points  to an $N$-dimensional spherical core of radius 0.3. In CHUNC2, all points are forced to a Gaussian distribution in the original dimensions, while the additional categorical quantity is constrained to a value of 1.0 (0.0) for valid (invalid) points.

Despite not using dimensional reduction, CHUNC and CHUNC2 both achieve more than two orders of magnitude improvements in generation efficiency over the naive scan, see Table~\ref{tab:chunc}.
 In the {\cmssm}, CHUNC2 has an efficiency of 0.32, outperforming CHUNC's efficiency of 0.21. However, in the higher dimensional {\pmssm} case CHUNC's efficiency of 0.24 exceeds CHUNC2's efficiency of 0.15. It is difficult to say why this is the case. Iterative KDE sampling may improve these results further. 
 
\begin{figure}[H]
\centering
\includegraphics[width=\linewidth]{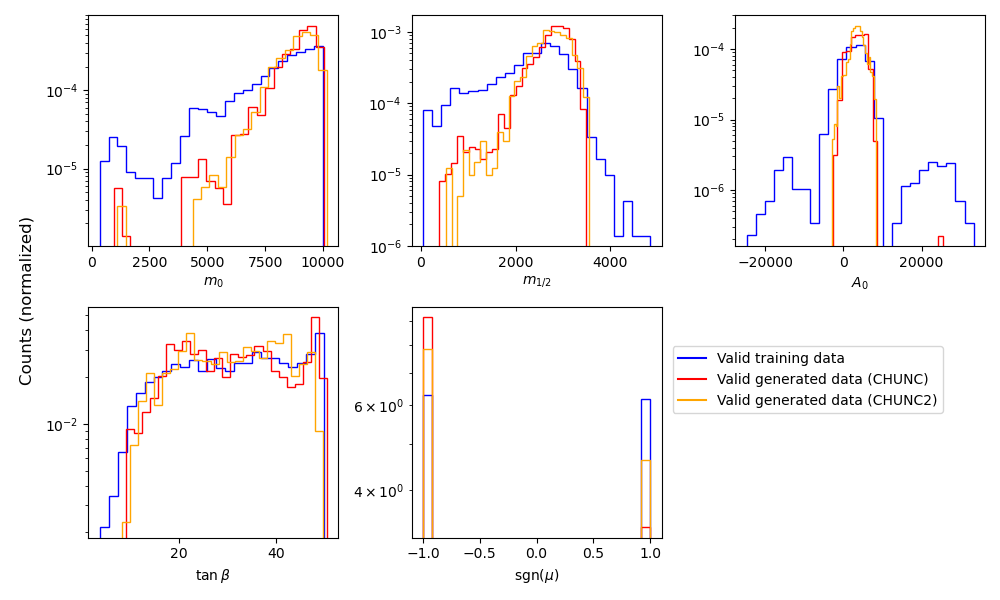}
\caption{ Distribution of CHUNC (red) and CHUNC2 generated (orange) points in each of the five {\cmssm} parameters under the triple constraint, compared to the distribution of the training dataset (blue). }
\label{fig:CHUNC visualization cmssm}
\end{figure}

\begin{figure}[H]
\centering
\includegraphics[width=\linewidth]{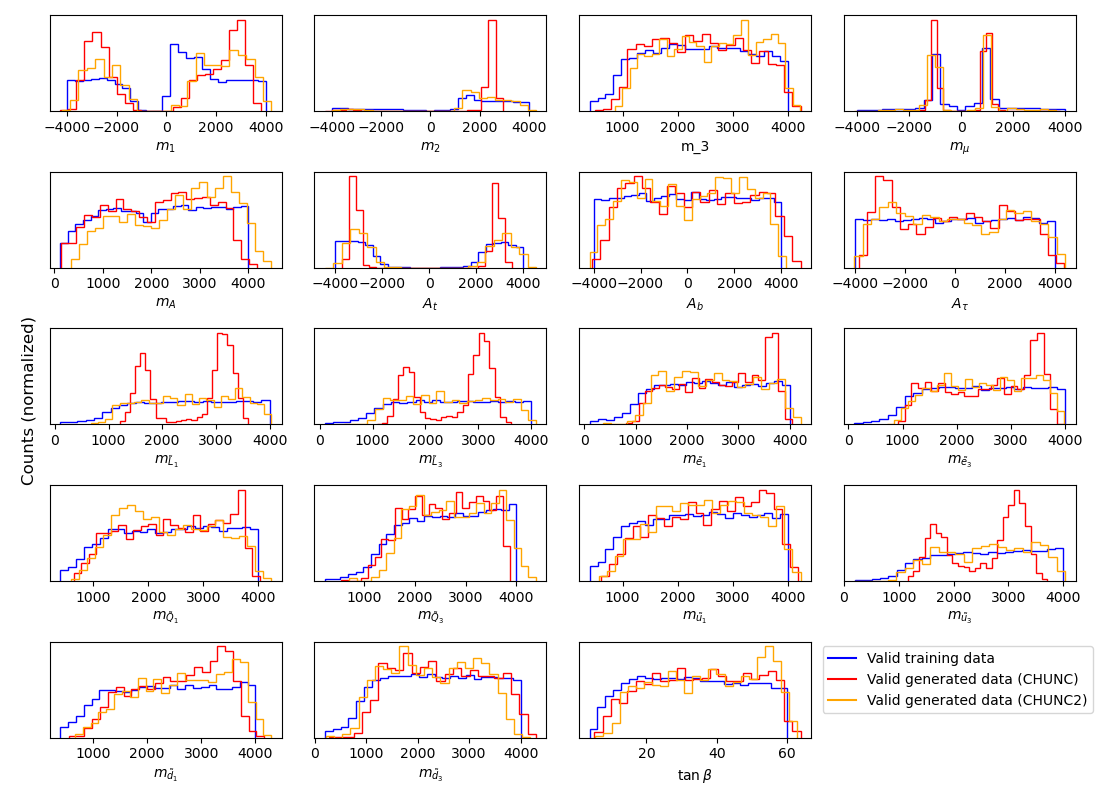}
\caption{Distribution of CHUNC (red) and CHUNC2 generated (orange) points in each of the nineteen {\pmssm} parameters under the triple constraint, compared to the distribution of the training dataset (blue).}
\label{fig:CHUNC visualization pmssm}
\end{figure}

 While CHUNC and CHUNC2 appear evenly matched at preserving the proportionality of the valid subspace in the lower dimensional {\cmssm} trials (see Fig.~\ref{fig:CHUNC visualization cmssm}),  CHUNC2 does a visibly better job at capturing the proportionality of the valid subspace without introducing significant distortion, as seen most notably in the $|M_2|$ , $|A_t|$ , $m_{\tilde{L}_1}$,$m_{\tilde{L}_3}$ 
and $m_{\tilde{u}_3}$ distributions in Fig.~\ref{fig:CHUNC visualization pmssm}. However, CHUNC outpeforms CHUNC2 for {\pmssm} efficiency. However our 'incompleteness' metric suggests these differences may be less prominent than they appear. Both CHUNC and CHUNC2 appear to introduce notably less bias than SAFESPAM when sampling from the {\pmssm}.

\begin{table}[h]
\caption{Comparison of the efficiency of generation of experimentally-valid points in naive scanning versus CHUNC and CHUNC2 methods for three experimental constraints (Higgs mass $m_H$, dark matter relic density $\Omega_{DM}$, and the lightest supersymmetric particle (LSP) being a neutralino) and two theoretical spaces (5-dim {\cmssm}, 19-dim {\pmssm}). The uncertainty across all trials is  $\sim\mathcal{O}(0.01)$, estimated by running 10 independent samplings for each case.}
\label{tab:chunc}
\centering
\begin{tabular}{c |c| c c l} 
 \hline \hline
 & &\multicolumn{3}{c}{Method}\\
  Space & Constraint & Naive scan &  CHUNC  & CHUNC2 \\ [0.5ex] 
  \hline
 {\cmssm} & $m_H \cap \Omega_{DM} \cap$ LSP & 0.0004 & 0.21 & 0.32  \\ 
 {\pmssm} & $m_H \cap \Omega_{DM} \cap$ LSP & 0.0012 & 0.24 & 0.15   \\ 
\hline
 \hline
\end{tabular}
\end{table}

\begin{table}[H]
\centering
\caption{ {\color{blue}Comparison of the Core-Forming and incompleteness Metrics for  CHUNC and CHUNC2  for the 'triple' experimental constraint (Higgs mass $m_H$ $\cap$ dark matter relic density $\Omega_{DM}$ $cap$ Neutralino LSP,) in two theoretical spaces (5-dim {\cmssm}, 19-dim {\pmssm}). Note that CHUNC2's categorical variable is excluded from calculations of the Core-Forming metric.}}
\label{table:eff3}
{\color{blue} \begin{tabular}{c |c|c| c c c } 
 \hline \hline
  Space & Method & Constraint & Core-Forming &  Incompleteness \\ [0.5ex] 
 \hline
 {\cmssm} & CHUNC & $m_H \cap \Omega_{DM} \cap$ LSP & 7.9545 & 0.0252 \\ 
 {\pmssm} & CHUNC & $m_H \cap \Omega_{DM} \cap$ LSP &  0.0212 & 0.0023    \\   
 \hline
 {\cmssm} & CHUNC2 & $m_H \cap \Omega_{DM} \cap$ LSP & 0.0163 & 0.0174\\
 {\pmssm} & CHUNC2 & $m_H \cap \Omega_{DM} \cap$ LSP & 0.0007 & 0.0020 \\
\hline
\end{tabular}}
\end{table}

\section{Discussion and Conclusion}
\subsection{Strengths and Weaknesses }

SAFESPAM is capable of  significant improvements in efficiency over naive sampling, attaining  greater than 95\% efficiency in all single-constraint cases; see Table~\ref{table:eff}. However, the range of points sampled often missed valid regions from the training data, especially in the {\pmssm}. In many cases we also found valid points  generated outside the boundaries specified in our initial data generation, in ranges previously unknown to the model, behavior not unexpected for such models. For example,  Fig. ~\ref{fig:Higgs cmssm} shows valid points with $m_0, m_{1/2} > 10,000$ GeV. While the latent distributions produced by SAFESPAM often display a distinct boundary between valid and invalid points, this was not the case for the Dark Matter trials, perhaps due to the narrower constraint in weak-scale space.  

{\color{blue} Differences in evaluation software, differing combinations of experimental constraints, variations on which parameters of the spaces are being explored and the range of values over which they are explored all make it a significant challenge to make direct comparisons to other state of the art methods. In most cases these differences constitute essentially different though related problems with potentially significant differences in level of difficulty. That said, if we ignore this we find that for the Higgs mass constraint, in both the {\cmssm} and {\pmssm} SAFESPAM achieves efficiencies which significantly outperforms prior work using HMC (0.723 {\cmssm} and 0.319 {\pmssm}), normalizing flows (0.796 {\cmssm} and 0.663 {\pmssm}), NSGA-II (0.715 {\cmssm} and 0.862 {\pmssm})and TPE (0.668 {\cmssm} and 0.557 {\pmssm}) methods, and slightly outperforms CMA-ES (0.924 {\cmssm} and 0.899 {\pmssm})~\cite{Hollingsworth:2021sii, Souza}. A significant caveat is that SAFESPAM often fails to capture portions of the valid subspace, especially in the PMSSM. Other methods do not seem to struggle with this to the same extent (although this is difficult to quantify since few papers provide a numerical metric which directly measures this sort of sampling bias in aggregate over all dimensions). 

\par
Across SAFESPAM, CHUNC, and CHUNC2 individually, we see that better (lower core-forming) values seem to generally correspond with better (lower incompleteness), but comparing across methods suggests that the removal of the constraints of dimensional reduction allows for greater leeway in this regard: In SAFESPAM core-formation seems to become more difficult as the number of dimensions increases, but for CHUNC and CHUNC2 the opposite holds true. Although CHUNC seems to struggle with core formation at times even more than SAFESPAM, this does not seem to effect its incompleteness nearly as much, perhaps because of the method's lack of dimensional reduction. Notably in the {\cmssm} CHUNC has a core-forming value several orders of magnitude worse than any other trial, but is still able to achieve better incompleteness than our {\pmssm} SAFESPAM dark matter trial. With respect to core-forming alone, CHUNC2 seems to perform around 2 orders of magnitude better than corresponding CHUNC trials in both the {\pmssm} and {\cmssm} respectively, perhaps because the contrastive task has been shifted onto the categorical variable. However it is interesting and worthwhile to note the lack of pronounced effect this core-forming issue has on incompleteness in CHUNC and CHUNC2, compared to SAFESPAM. It is also possible that the differences in incompleteness we observe are due to CHUNC and CHUNC2 being trained on a combination of valid and invalid points where SAFESPAM was trained only on valid data.  Additionally we see that although CHUNC2 outperforms CHUNC in both the core-forming and incompleteness metrics in both spaces, this does not always lead to better efficiency. It is difficult to quantify definitively why this is the case.}

Both SAFESPAM and CHUNC-style methods show notable improvement across several constraints compared to naive sampling in the {\cmssm} and {\pmssm}. The ability to apply additional constraints while sampling makes these maps versatile tools, and the iterative usage of KDE sampling guarantees that even difficult latent shapes can be sampled with high efficiency.  Optimistically, one imagines a web of theorists training SWAE networks on specific sub-regions of interest in these parameter spaces and sharing their models with one another for easy data generation in those regions. Another possibility is to use these methods to explore lesser-known and under-explored subspaces within the full MSSM, such as those which might lead to new solutions to the little hierarchy problem ~\cite{feruglio2023fermion,barbieri2000lep}. Our methodology could also be used with non-MSSM theoretical parameter spaces, or any high-dimensional supervised learning problem with a clear labeled distinction between valid and invalid points. 

\par Broadly, there is a clear trade-off between the efficiency benefits of dimensional reduction methods such as SAFESPAM and the potential sampling biases introduced by compressing these high-dimensional spaces.  In contrast, CHUNC and CHUNC2-style networks avoid dimensional reduction in their latent spaces, and achieve significantly less bias in sampling the {\pmssm} despite imposing no constraints that select for this, allowing these methods to sample a broader range of valid points. However, with no dimensional reduction, these latent spaces are still vast and difficult to search exhaustively, even if well-organized, and KDE-style sampling can be less effective, resulting in lower yield.

\subsection{Future Avenues}

In terms of near term improvements of our methodology, sampling jointly from an ensemble of models with varying hyperparameters could help recover portions of these spaces lost due to dimensional reduction in any given individual model. Alternatively, one might introduce a secondary Sliced Wasserstein loss term, applied to fit randomly generated output data to the expected proportionality of the valid sub-space while training, in order to prevent biased sampling.

\par
The {\cmssm} and {\pmssm} both adopt de facto theory biases in constraining the underlying MSSM space. Ideally, machine-learning-driven methods will someday be powerful enough to  tackle larger unconstrained spaces with high accuracy and efficiency, usurping the need for such constraints. However neither the SAFESPAM or CHUNC-style method in their current iterations would perform well in a $\sim$100 parameter space such as the unconstrained MSSM. Reduction to two dimensions via a SAFESPAM-style methodology would cause significant information loss, and a CHUNC-style method with no dimensional reduction would be overwhelmed by the sheer vastness of the space, the primary reason the MSSM is difficult to explore in the first place. Standard KDE sampling is also known to degrade due to curse of dimensionality, though there are proposed methods to circumvent this ~\cite{Nagler_2016,10.1007/978-3-642-60126-2_36,https://doi.org/10.48550/arxiv.2212.00759}. Ultimately, future work in this vein will need to thread the needle with dimensional reduction in order to simplify sampling and search methods without incurring significant information loss. One optimistically imagines a variable-dimension auto-encoder where the dimension of the latent space at any given point is itself a trainable parameter. Such a network could constrain different portions of a space into separate yet interconnected lower-dimensional representations, changing the dimension of the latent space as needed to maximize dimensional reduction benefits while avoiding information loss. The boundaries of the resulting structures in the latent space could be elucidated iteratively then those structures could be learned and sampled via manifold learning ~\cite{park2022neural,brehmer2020flows}. \par

Another fundamental issue is that as dimensionality increases a larger initial scan is needed to represent the space without losing resolution. In the unconstrained MSSM any sort of uniform scan is currently intractable: a scan with only 2 values per dimension would require $\sim 2^{100}$ points. A potential solution to this would be to focus on sequential or algorithmic search methods that can operate without the overhead of evaluating a large initial dataset  (see ~\cite{Morrison:2022},~\cite{https://doi.org/10.48550/arxiv.1805.03615} for examples), but it is unclear how exactly one would meld these with a dimensional-reduction-based framework. 
    
\subsection{Conclusion}   
 We trained a variety of modified Sliced Wasserstein autoencoders, both with and without dimensional reduction, to sample efficiently from experimentally valid regions of two sub-spaces of the MSSM, the {\cmssm} and {\pmssm}, under multiple experimental constraints. Utilizing these mappings alongside KDE sampling, we were able to demonstrate efficiencies significantly greater than naive values in all cases, and we suspect additional iterative usage of KDE sampling could improve these results even further. Our method also is capable of easily adding additional secondary constraints without any retraining, in part due to the benefits of dimensional reduction. 
 \par
 If subtle, unexplored regions remain in spaces like the {\cmssm} and {\pmssm}, their discovery hinges on sampling methods that can overcome the inherent difficulties of searching high-dimensional spaces. Methods like SAFESPAM, CHUNC, and CHUNC2 provide a novel way to drill deeper into such parameter spaces and ensure underexplored regions do not go overlooked. Further work with these methods to analyze such regions in the {\cmssm} and {\pmssm} remains, as do additional iterative strategies to apply additional successive experimental constraints. The discovery of notable regions in these spaces could easily inspire to new LHC analyses in search of supersymmetric signatures we would not know to look for otherwise.

\acknowledgments

Special thanks to Aishik Ghosh and Jessica N. Howard for their valuable advice and discussion, as well as Max Fieg, Kevin Greif, Gregor Kasieczka, Michael Ratz, and David Shih for their feedback and perspective.

\appendix
\section{Hyperparameters}
Hyperparameters for SAFESPAM, CHUNC and CHUNC2 

\begin{table}[H]
\centering
\caption{ Table of Hyperparameters for SAFESPAM. Note for pMSSM $m_H$ trial 1 training was split between 20 epochs with a proportional mix of valid and invalid data and 20 epochs of only valid data.}
\label{table:hyperparams1}
\begin{tabular}{c | c c c c c } 
 \hline \hline
  Space & cMSSM  & cMSSM  & pMSSM  & pMSSM  & pMSSM  \\ [0.5ex] 
  \hline
Constraint & $m_H$ & $\Omega_{DM}$ & $m_H$ (Tr. 1) & $m_H$ (Tr. 2) & $\Omega_{DM}$ \\
 \hline
 Batch Sz. & 2500 & 2500 & 2500 & 2500 & 2500  \\ 
 L.R. & 0.0005 & 0.0005 & 0.0005 & 0.0005 & 0.0005 \\
 Opt. & ADAM & ADAM & ADAM & ADAM & ADAM\\  
 num. layers & 10 & 10 & 10 & 10 & 12  \\  
 Act. & Leaky RELU & Leaky RELU & Leaky RELU & Leaky RELU & Leaky RELU \\
 num. epochs & 60 & 6160 & 40*  & 160 & 160  \\ 
 $\alpha$  & 5 & 5 & 5 & 5 & 5  \\
 $\beta$  & 1 & 5 & 1 & 1 & 1 \\
 $\gamma$ & 0 & 0.1 & 0.3 & 0.3 & 0.8 \\
\hline \hline
\end{tabular}
\end{table}

\begin{table}[H]
\centering
\caption{ Table of Hyperparameters for CHUNC and CHUNC2. }
\label{table:hyperparams2}
\begin{tabular}{c | c c c c  } 
 \hline \hline
 Model & CHUNC & CHUNC2 & CHUNC & CHUNC2 \\
 \hline
 Space & cMSSM  & cMSSM  & pMSSM  & pMSSM    \\ [0.5ex] 
  \hline
Constraint & Triple & Triple & Triple & Triple \\
 \hline
 Batch Sz. & 64 & 64 & 64 & 64    \\ 
 L.R. & 0.001 & 0.001 & 0.001 & 0.001  \\
 Opt. & ADAM & ADAM & ADAM & ADAM \\  
 num. layers & 22 & 22 & 22 & 22  \\  
 Act. & Leaky RELU & Leaky RELU & Leaky RELU & Leaky RELU  \\
 num. epochs & 100 & 100 & 100  & 100   \\ 
 $\alpha$  & 1 & 1 & 1 & 1   \\
 $\beta$  & 1 & 1 & 1 & 1  \\
 $\gamma$ & 1 & 1 & 1 & 1  \\
 c & 1 & N/A & 1 & N/A  \\
 d & 1 & N/A & 1 & N/A \\
\hline \hline
\end{tabular}
\end{table}

\bibliographystyle{JHEP}
\bibliography{ref} 

\providecommand{\href}[2]{#2}\begingroup\raggedright\begin{thebibliography}{10}

\bibitem{MARTIN_1998}
S.~P. MARTIN, {\it A {SUPERSYMMETRY} {PRIMER}},  in {\em Perspectives on
  Supersymmetry}, pp.~1--98.
\newblock {WORLD} {SCIENTIFIC}, jul, 1998.

\bibitem{PhysRevD.86.055007}
D.~Ghosh, M.~Guchait, S.~Raychaudhuri, and D.~Sengupta, {\it How constrained is
  the constrained mssm?},  {\em Phys. Rev. D} {\bf 86} (Sep, 2012) 055007.

\bibitem{HAN2017470}
C.~Han, K.~ichi Hikasa, L.~Wu, J.~M. Yang, and Y.~Zhang, {\it Status of cmssm
  in light of current lhc run-2 and lux data},  {\em Physics Letters B} {\bf
  769} (2017) 470--476.

\bibitem{PhysRevD.83.095019}
B.~C. Allanach, {\it Impact of cms multi-jets and missing energy search on
  cmssm fits},  {\em Phys. Rev. D} {\bf 83} (May, 2011) 095019.

\bibitem{Buchmueller2014}
O.~Buchmueller, R.~Cavanaugh, A.~Roeck, and et~al., {\it {The CMSSM and NUHM1
  after LHC Run 1.}},  {\em Eur. Phys. J. C} {\bf 74} (2014), no.~2922.

\bibitem{Bridges:2011}
M.~Bridges, K.~Cranmer, and F.~Feroz~et al., {\it {A coverage study of the
  CMSSM based on ATLAS sensitivity using fast neural networks techniques}},
  {\em J. High Energ. Phys.} (2011), no.~12.

\bibitem{https://doi.org/10.48550/arxiv.1307.8444}
M.~Cahill-Rowley, J.~L. Hewett, A.~Ismail, and T.~G. Rizzo, {\it pmssm studies
  at the 7, 8 and 14 tev lhc},  2013.

\bibitem{PhysRevD.91.055002}
M.~Cahill-Rowley, J.~L. Hewett, A.~Ismail, and T.~G. Rizzo, {\it Lessons and
  prospects from the pmssm after lhc run i},  {\em Phys. Rev. D} {\bf 91} (Mar,
  2015) 055002.

\bibitem{Aad_2015}
G.~Aad, B.~Abbott, J.~Abdallah, O.~Abdinov, R.~Aben, M.~Abolins, O.~S.
  AbouZeid, H.~Abramowicz, H.~Abreu, and et~al., {\it Summary of the atlas
  experiment’s sensitivity to supersymmetry after lhc run 1 — interpreted
  in the phenomenological mssm},  {\em Journal of High Energy Physics} {\bf
  2015} (Oct, 2015).

\bibitem{CMS:2016}
T.~C. collaboration., V.~Khachatryan, A.~Sirunyan, and et~al., {\it
  {Phenomenological MSSM interpretation of CMS searches in pp collisions at
  $\sqrt{s} =$ 7 and 8 TeV}},  {\em J. High Energ. Phys.} (2016), no.~129.

\bibitem{Caron2017}
S.~Caron, J.~S. Kim, K.~Rolbiecki, R.~R. de~Austri, and B.~Stienen, {\it The
  bsm-ai project: Susy-ai--generalizing lhc limits on supersymmetry with
  machine learning},  {\em The European Physical Journal C} {\bf 77} (Apr,
  2017) 257.

\bibitem{Kronheim_2021}
B.~Kronheim, M.~Kuchera, H.~Prosper, and A.~Karbo, {\it Bayesian neural
  networks for fast susy predictions},  {\em Physics Letters B} {\bf 813} (Feb,
  2021) 136041.

\bibitem{Bridges_2011}
M.~Bridges, K.~Cranmer, F.~Feroz, M.~Hobson, R.~Ruiz~de Austri, and R.~Trotta,
  {\it A coverage study of the cmssm based on atlas sensitivity using fast
  neural networks techniques},  {\em Journal of High Energy Physics} {\bf 2011}
  (Mar, 2011).

\bibitem{Cheung:2012qy}
C.~Cheung, L.~J. Hall, D.~Pinner, and J.~T. Ruderman, {\it {Prospects and Blind
  Spots for Neutralino Dark Matter}},  {\em JHEP} {\bf 05} (2013) 100,
  [\href{https://arxiv.org/abs/1211.4873}{{\tt arXiv:1211.4873}}].

\bibitem{Feng:1999zg}
J.~L. Feng, K.~T. Matchev, and T.~Moroi, {\it {Focus points and naturalness in
  supersymmetry}},  {\em Phys. Rev. D} {\bf 61} (2000) 075005,
  [\href{https://arxiv.org/abs/hep-ph/9909334}{{\tt hep-ph/9909334}}].

\bibitem{Papucci:2011wy}
M.~Papucci, J.~T. Ruderman, and A.~Weiler, {\it {Natural SUSY Endures}},  {\em
  JHEP} {\bf 09} (2012) 035, [\href{https://arxiv.org/abs/1110.6926}{{\tt
  arXiv:1110.6926}}].

\bibitem{Dine_1993}
M.~Dine and A.~E. Nelson, {\it Dynamical supersymmetry breaking at low
  energies},  {\em Physical Review D} {\bf 48} (aug, 1993) 1277--1287.

\bibitem{Meade_2009}
P.~Meade, N.~Seiberg, and D.~Shih, {\it General gauge mediation},  {\em
  Progress of Theoretical Physics Supplement} {\bf 177} (2009) 143--158.

\bibitem{Kowalska2013NaturalMA}
K.~Kowalska and E.~M. Sessolo, {\it Natural mssm after the lhc 8 tev run},
  {\em Physical Review D} {\bf 88} (2013) 075001.

\bibitem{Hall2011ANS}
L.~J. Hall, D.~Pinner, and J.~T. Ruderman, {\it A natural susy higgs near 125
  gev},  {\em Journal of High Energy Physics} {\bf 2012} (2011) 1--25.

\bibitem{Huang2014BlindSF}
P.~Huang and C.~E.~M. Wagner, {\it Blind spots for neutralino dark matter in
  the mssm with an intermediate m\_a},  {\em Physical Review D} {\bf 90} (2014)
  015018.

\bibitem{Feng_2012}
J.~L. Feng and D.~Sanford, {\it Natural 125 gev higgs boson in the mssm from
  focus point supersymmetry with $a$-terms},  {\em Phys. Rev. D} {\bf 86} (Sep,
  2012) 055015.

\bibitem{Kowalska2014LowFT}
K.~Kowalska, L.~Roszkowski, E.~M. Sessolo, and S.~Trojanowski, {\it Low fine
  tuning in the mssm with higgsino dark matter and unification constraints},
  {\em Journal of High Energy Physics} {\bf 2014} (2014) 1--36.

\bibitem{Evans_2014}
J.~A. Evans, Y.~Kats, D.~Shih, and M.~J. Strassler, {\it Toward full {LHC}
  coverage of natural supersymmetry},  {\em Journal of High Energy Physics}
  {\bf 2014} (jul, 2014).

\bibitem{Buckley_2017_1}
M.~R. Buckley, D.~Feld, S.~Macaluso, A.~Monteux, and D.~Shih, {\it Cornering
  natural {SUSY} at {LHC} run {II} and beyond},  {\em Journal of High Energy
  Physics} {\bf 2017} (aug, 2017).

\bibitem{Buckley_2017_2}
M.~R. Buckley, A.~Monteux, and D.~Shih, {\it Precision corrections to fine
  tuning in {SUSY}},  {\em Journal of High Energy Physics} {\bf 2017} (jun,
  2017).

\bibitem{Dine_1995}
M.~Dine, A.~E. Nelson, and Y.~Shirman, {\it Low energy dynamical supersymmetry
  breaking simplified},  {\em Physical Review D} {\bf 51} (feb, 1995)
  1362--1370.

\bibitem{Dine_1996}
M.~Dine, A.~E. Nelson, Y.~Nir, and Y.~Shirman, {\it New tools for low energy
  dynamical supersymmetry breaking},  {\em Physical Review D} {\bf 53} (mar,
  1996) 2658--2669.

\bibitem{goodfellow2014generative}
I.~J. Goodfellow, J.~Pouget-Abadie, M.~Mirza, B.~Xu, D.~Warde-Farley, S.~Ozair,
  A.~Courville, and Y.~Bengio, {\it Generative adversarial networks},  2014.

\bibitem{betancourt2018conceptual}
M.~Betancourt, {\it A conceptual introduction to hamiltonian monte carlo},
  2018.

\bibitem{neal2012mcmc}
R.~M. Neal, {\it Mcmc using hamiltonian dynamics},  2012.

\bibitem{Baltz_2004}
E.~A. Baltz and P.~Gondolo, {\it Markov chain monte carlo exploration of
  minimal supergravity with implications for dark matter},  {\em Journal of
  High Energy Physics} {\bf 2004} (Oct, 2004) 052–052.

\bibitem{Kobyzev_2021}
I.~Kobyzev, S.~J. Prince, and M.~A. Brubaker, {\it Normalizing flows: An
  introduction and review of current methods},  {\em {IEEE} Transactions on
  Pattern Analysis and Machine Intelligence} {\bf 43} (nov, 2021) 3964--3979.

\bibitem{Goldberg1988GeneticAI}
D.~E. Goldberg, {\it Genetic algorithms in search optimization and machine
  learning},  1988.

\bibitem{Hollingsworth:2021sii}
J.~Hollingsworth, M.~Ratz, P.~Tanedo, and D.~Whiteson, {\it Efficient sampling
  of constrained high-dimensional theoretical spaces with machine learning},
  {\em The European Physical Journal C} {\bf 81} (dec, 2021).

\bibitem{Morrison:2022}
L.~Morrison, S.~Profumo, and J.~Tamanas, {\it Simulation based inference for
  efficient theory space sampling: an application to supersymmetric
  explanations of the anomalous muon (g-2)},  2022.

\bibitem{Abel_2014}
S.~Abel and J.~Rizos, {\it Genetic algorithms and the search for viable string
  vacua},  {\em Journal of High Energy Physics} {\bf 2014} (aug, 2014).

\bibitem{Souza}
F.~A. de~Souza, M.~C. Romão, N.~F. Castro, M.~Nikjoo, and W.~Porod, {\it
  Exploring parameter spaces with artificial intelligence and machine learning
  black-box optimisation algorithms},  2022.

\bibitem{https://doi.org/10.48550/arxiv.2207.05103}
J.~Dickinson, S.~Bein, S.~Heinemeyer, J.~Hiltbrand, J.~Hirschauer, W.~Hopkins,
  E.~Lipeles, M.~Mrowietz, and N.~Strobbe, {\it A grand scan of the pmssm
  parameter space for snowmass 2021},  2022.

\bibitem{https://doi.org/10.48550/arxiv.1805.03615}
S.~Abel, D.~G. Cerdeno, and S.~Robles, {\it The power of genetic algorithms:
  what remains of the pmssm?},  2018.

\bibitem{https://doi.org/10.48550/arxiv.1802.03426}
L.~McInnes, J.~Healy, and J.~Melville, {\it Umap: Uniform manifold
  approximation and projection for dimension reduction},  2018.

\bibitem{JMLR:v9:vandermaaten08a}
L.~van~der Maaten and G.~Hinton, {\it Visualizing data using t-sne},  {\em
  Journal of Machine Learning Research} {\bf 9} (2008), no.~86 2579--2605.

\bibitem{ACKLEY1985147}
D.~H. Ackley, G.~E. Hinton, and T.~J. Sejnowski, {\it A learning algorithm for
  boltzmann machines},  {\em Cognitive Science} {\bf 9} (1985), no.~1 147--169.

\bibitem{2007SchpJ...2.1568K}
T.~{Kohonen} and T.~{Honkela}, {\it {Kohonen network}},  {\em Scholarpedia}
  {\bf 2} (Jan., 2007) 1568.

\bibitem{Kohonen2004SelforganizedFO}
T.~Kohonen, {\it Self-organized formation of topologically correct feature
  maps},  {\em Biological Cybernetics} {\bf 43} (2004) 59--69.

\bibitem{Mutter:2019}
A.~Mütter, E.~Parr, and P.~K. Vaudrevange, {\it Deep learning in the heterotic
  orbifold landscape},  {\em Nuclear Physics B} {\bf 940} (mar, 2019) 113--129.

\bibitem{he2022machinelearning}
Y.-H. He and J.~M.~P. Ipiña, {\it Machine-learning the classification of
  spacetimes},  2022.

\bibitem{Fefferman_Mitter_Narayanan_2016}
C.~Fefferman, S.~Mitter, and H.~Narayanan, {\it Testing the manifold
  hypothesis},  {\em Journal of the American Mathematical Society} {\bf 29}
  (Oct, 2016) 983–1049.

\bibitem{Kolouri:2018}
S.~Kolouri, P.~E. Pope, C.~E. Martin, and G.~K. Rohde, {\it Sliced-wasserstein
  autoencoder: An embarrassingly simple generative model},  2018.

\bibitem{park2022neural}
S.~E. Park, P.~Harris, and B.~Ostdiek, {\it Neural embedding: Learning the
  embedding of the manifold of physics data},  2022.

\bibitem{brehmer2020flows}
J.~Brehmer and K.~Cranmer, {\it Flows for simultaneous manifold learning and
  density estimation},  2020.

\bibitem{kumar2022manifold}
A.~Kumar and M.~Sarovar, {\it Manifold learning via quantum dynamics},  2022.

\bibitem{kumar2022shining}
A.~Kumar and M.~Sarovar, {\it Shining light on data: Geometric data analysis
  through quantum dynamics},  2022.

\bibitem{CHUNCNet}
N.~Carrara, ``Constraint-driven high-dimensional uncompressed (categorical)
  clustering (chuncnet).'' \url{https://github.com/infophysics/CHUNCNet}, 2023.

\bibitem{Cohen:2013}
T.~Cohen and J.~G. Wacker, {\it Here be dragons: the unexplored continents of
  the {CMSSM}},  {\em Journal of High Energy Physics} {\bf 2013} (sep, 2013).

\bibitem{Djouadi:1999}
A.~Djouadi, S.~Rosier-Lees, M.~Bezouh, M.~A. Bizouard, C.~Boehm, F.~Borzumati,
  C.~Briot, J.~Carr, M.~B. Causse, F.~Charles, X.~Chereau, P.~Colas, L.~Duflot,
  A.~Dupperin, A.~Ealet, H.~El-Mamouni, N.~Ghodbane, F.~Gieres,
  B.~Gonzalez-Pineiro, S.~Gourmelen, G.~Grenier, P.~Gris, J.~F. Grivaz,
  C.~Hebrard, B.~Ille, J.~L. Kneur, N.~Kostantinidis, J.~Layssac, P.~Lebrun,
  R.~Ledu, M.~C. Lemaire, C.~LeMouel, L.~Lugnier, Y.~Mambrini, J.~P. Martin,
  G.~Montarou, G.~Moultaka, S.~Muanza, E.~Nuss, E.~Perez, F.~M. Renard,
  D.~Reynaud, L.~Serin, C.~Thevenet, A.~Trabelsi, F.~Zach, and D.~Zerwas, {\it
  The minimal supersymmetric standard model: Group summary report},  1999.

\bibitem{PhysRevLett.49.970}
A.~H. Chamseddine, R.~Arnowitt, and P.~Nath, {\it Locally supersymmetric grand
  unification},  {\em Phys. Rev. Lett.} {\bf 49} (Oct, 1982) 970--974.

\bibitem{20121}
G.~Aad, T.~Abajyan, B.~Abbott, J.~Abdallah, S.~{Abdel Khalek}, A.~Abdelalim,
  O.~Abdinov, R.~Aben, B.~Abi, M.~Abolins, and et~al., {\it Observation of a
  new particle in the search for the standard model higgs boson with the atlas
  detector at the lhc},  {\em Physics Letters B} {\bf 716} (2012), no.~1 1--29.

\bibitem{ce431d39f29d462482f95f2cb27c9583}
S.~Chatrchyan, V.~Khachatryan, A.~Sirunyan, A.~Tumasyan, W.~Adam, E.~Aguilo,
  T.~Bergauer, M.~Dragicevic, J.~Er{\"o}, C.~Fabjan, and et~al., {\it Study of
  the mass and spin-parity of the higgs boson candidate via its decays to z
  boson pairs},  {\em Physical Review Letters} {\bf 110} (Feb., 2013).

\bibitem{Spergel_2003}
D.~N. Spergel, L.~Verde, H.~V. Peiris, E.~Komatsu, M.~R. Nolta, C.~L. Bennett,
  M.~Halpern, G.~Hinshaw, N.~Jarosik, A.~Kogut, and et~al., {\it First‐year
  wilkinson microwave anisotropy probe ( wmap ) observations: Determination of
  cosmological parameters},  {\em The Astrophysical Journal Supplement Series}
  {\bf 148} (Sep, 2003) 175–194.

\bibitem{Bennett_2013}
C.~L. Bennett, D.~Larson, J.~L. Weiland, N.~Jarosik, G.~Hinshaw, N.~Odegard,
  K.~M. Smith, R.~S. Hill, B.~Gold, M.~Halpern, E.~Komatsu, M.~R. Nolta,
  L.~Page, D.~N. Spergel, E.~Wollack, J.~Dunkley, A.~Kogut, M.~Limon, S.~S.
  Meyer, G.~S. Tucker, and E.~L. Wright, {\it {NINE}-{YEAR} {WILKINSON}
  {MICROWAVE} {ANISOTROPY} {PROBE} ( {WMAP} ) {OBSERVATIONS}: {FINAL} {MAPS}
  {AND} {RESULTS}},  {\em The Astrophysical Journal Supplement Series} {\bf
  208} (sep, 2013) 20.

\bibitem{ALLANACH2002305}
B.~Allanach, {\it Softsusy: A program for calculating supersymmetric spectra},
  {\em Computer Physics Communications} {\bf 143} (2002), no.~3 305--331.

\bibitem{micrOMEGAs}
G.~Belanger, F.~Boudjema, A.~Pukhov, and A.~Semenov, {\it micromegas: A tool
  for dark matter studies},  {\em Il Nuovo Cimento C} {\bf 33} (Aug, 2010)
  111–116.

\bibitem{VAE}
D.~P. Kingma and M.~Welling, {\it Auto-encoding variational bayes},  2013.

\bibitem{Tolstikhin:2017}
I.~Tolstikhin, O.~Bousquet, S.~Gelly, and B.~Schoelkopf, {\it Wasserstein
  auto-encoders},  2019.

\bibitem{Howard:2021pos}
J.~N. Howard, S.~Mandt, D.~Whiteson, and Y.~Yang, {\it Learning to simulate
  high energy particle collisions from unlabeled data},  {\em Scientific
  Reports} {\bf 12} (may, 2022).

\bibitem{zheng2018degeneration}
H.~Zheng, J.~Yao, Y.~Zhang, and I.~W. Tsang, {\it Degeneration in vae: in the
  light of fisher information loss},  2018.

\bibitem{NEURIPS2019_9015}
A.~Paszke, S.~Gross, F.~Massa, A.~Lerer, J.~Bradbury, G.~Chanan, T.~Killeen,
  Z.~Lin, N.~Gimelshein, L.~Antiga, A.~Desmaison, A.~Kopf, E.~Yang, Z.~DeVito,
  M.~Raison, A.~Tejani, S.~Chilamkurthy, B.~Steiner, L.~Fang, J.~Bai, and
  S.~Chintala, {\it Pytorch: An imperative style, high-performance deep
  learning library},  in {\em Advances in Neural Information Processing Systems
  32} (H.~Wallach, H.~Larochelle, A.~Beygelzimer, F.~d\textquotesingle
  Alch\'{e}-Buc, E.~Fox, and R.~Garnett, eds.), pp.~8024--8035.
\newblock Curran Associates, Inc., 2019.

\bibitem{ADAM}
D.~P. Kingma and J.~Ba, {\it Adam: A method for stochastic optimization},
  2014.

\bibitem{MLS}
D.~Kirkby, ``Machine learning statistics.''
  \url{https://github.com/dkirkby/MachineLearningStatistics}, 2021.

\bibitem{torchSWAE}
E.~Fuentes, ``Sliced-wasserstein autoencoder - pytorch.''
  \url{https://github.com/eifuentes/swae-pytorch}, 2018.

\bibitem{10.5555/1593511}
G.~Van~Rossum and F.~L. Drake, {\em Python 3 Reference Manual}.
\newblock CreateSpace, Scotts Valley, CA, 2009.

\bibitem{2020NumPy-Array}
C.~R. Harris, K.~J. Millman, S.~J. van~der Walt, R.~Gommers, P.~Virtanen,
  D.~Cournapeau, E.~Wieser, J.~Taylor, S.~Berg, N.~J. Smith, R.~Kern, M.~Picus,
  S.~Hoyer, M.~H. van Kerkwijk, M.~Brett, A.~Haldane, J.~Fernández~del Río,
  M.~Wiebe, P.~Peterson, P.~Gérard-Marchant, K.~Sheppard, T.~Reddy,
  W.~Weckesser, H.~Abbasi, C.~Gohlke, and T.~E. Oliphant, {\it Array
  programming with {NumPy}},  {\em Nature} {\bf 585} (2020) 357–362.

\bibitem{mckinney2010data}
W.~McKinney et~al., {\it Data structures for statistical computing in python},
  in {\em Proceedings of the 9th Python in Science Conference}, vol.~445,
  pp.~51--56, Austin, TX, 2010.

\bibitem{Waskom2021}
M.~L. Waskom, {\it seaborn: statistical data visualization},  {\em Journal of
  Open Source Software} {\bf 6} (2021), no.~60 3021.

\bibitem{hunter2007matplotlib}
J.~D. Hunter, {\it Matplotlib: A 2d graphics environment},  {\em Computing in
  science \& engineering} {\bf 9} (2007), no.~3 90--95.

\bibitem{Schroff:2015}
F.~Schroff, D.~Kalenichenko, and J.~Philbin, {\it {FaceNet}: A unified
  embedding for face recognition and clustering},  in {\em 2015 {IEEE}
  Conference on Computer Vision and Pattern Recognition ({CVPR})}, {IEEE}, jun,
  2015.

\bibitem{NIPS2003_d3b1fb02}
M.~Schultz and T.~Joachims, {\it Learning a distance metric from relative
  comparisons},  in {\em Advances in Neural Information Processing Systems}
  (S.~Thrun, L.~Saul, and B.~Sch\"{o}lkopf, eds.), vol.~16, MIT Press, 2003.

\bibitem{scikit-learn}
F.~Pedregosa, G.~Varoquaux, A.~Gramfort, V.~Michel, B.~Thirion, O.~Grisel,
  M.~Blondel, P.~Prettenhofer, R.~Weiss, V.~Dubourg, J.~Vanderplas, A.~Passos,
  D.~Cournapeau, M.~Brucher, M.~Perrot, and E.~Duchesnay, {\it Scikit-learn:
  Machine learning in {P}ython},  {\em Journal of Machine Learning Research}
  {\bf 12} (2011) 2825--2830.

\bibitem{CoverThomas}
T.~M. Cover and J.~A. Thomas, {\em Elements of Information Theory (Wiley Series
  in Telecommunications and Signal Processing)}.
\newblock Wiley-Interscience, USA, 2006.

\bibitem{feruglio2023fermion}
F.~Feruglio, {\it Fermion masses, critical behavior and universality},  2023.

\bibitem{barbieri2000lep}
R.~Barbieri and A.~Strumia, {\it The `lep paradox'},  2000.

\bibitem{Nagler_2016}
T.~Nagler and C.~Czado, {\it Evading the curse of dimensionality in
  nonparametric density estimation with simplified vine copulas},  {\em Journal
  of Multivariate Analysis} {\bf 151} (oct, 2016) 69--89.

\bibitem{10.1007/978-3-642-60126-2_36}
M.~Di~Marzio and G.~Lafratta, {\it Reducing dimensionality effects on kernel
  density estimation: The bivariate gaussian case},  in {\em Classification and
  Data Analysis} (M.~Vichi and O.~Opitz, eds.), (Berlin, Heidelberg),
  pp.~287--294, Springer Berlin Heidelberg, 1999.

\bibitem{https://doi.org/10.48550/arxiv.2212.00759}
Y.~Ren, H.~Zhao, Y.~Khoo, and L.~Ying, {\it High-dimensional density estimation
  with tensorizing flow},  2022.

\end{thebibliography}\endgroup

\end{document}